\documentclass[aps,prb,twocolumn,superscriptaddress,floatfix]{revtex4-1}

\usepackage{graphicx}
\usepackage{dcolumn}
\usepackage{bm}
\usepackage{amssymb}
\usepackage{microtype}
\usepackage{xfrac}
\usepackage{makecell}
\usepackage{array}
\usepackage[version=4]{mhchem}
\usepackage{gensymb}
\usepackage{multirow}
\usepackage[colorlinks=true,bookmarks=false]{hyperref}
\usepackage{comment}
\usepackage{chngcntr}

\newcommand{\angstrom}{\text{\normalfont\AA}}
\newcommand{\uvec}[1]{\boldsymbol{\hat{\textbf{#1}}}}

\begin{document}

\title{Anti-chiral spin order, its soft modes, and their hybridization with phonons\\ in the topological semimetal \ce{Mn3Ge}}

\author{Y.~Chen}
\affiliation{Institute for Quantum Matter and Department of Physics and Astronomy, Johns Hopkins University, Baltimore, MD 21218, USA}

\author{J.~Gaudet}
\affiliation{Institute for Quantum Matter and Department of Physics and Astronomy, Johns Hopkins University, Baltimore, MD 21218, USA}

\author{S.~Dasgupta}
\affiliation{Institute for Quantum Matter and Department of Physics and Astronomy, Johns Hopkins University, Baltimore, MD 21218, USA}

\author{G.~G.~Marcus}
\affiliation{Institute for Quantum Matter and Department of Physics and Astronomy, Johns Hopkins University, Baltimore, MD 21218, USA}

\author{J.~Lin}
\affiliation{Neutron Scattering Division, Oak Ridge National Laboratory, Oak Ridge, Tennessee 37831, USA}

\author{T.~Chen}
\affiliation{Institute for Solid State Physics (ISSP), University of Tokyo, Kashiwa, Chiba 277-8581, Japan}

\author{T.~Tomita}
\affiliation{Institute for Solid State Physics (ISSP), University of Tokyo, Kashiwa, Chiba 277-8581, Japan}

\author{M.~Ikhlas}
\affiliation{Institute for Solid State Physics (ISSP), University of Tokyo, Kashiwa, Chiba 277-8581, Japan}

\author{Y.~Zhao}
\affiliation{NIST Center for Neutron Research, National Institute of Standards and Technology, Gaithersburg, Maryland 20899, USA}
\affiliation{Department of Materials Science and Engineering, University of Maryland, College Park, Maryland 20742, USA}

\author{W.~C.~Chen}
\affiliation{NIST Center for Neutron Research, National Institute of Standards and Technology, Gaithersburg, Maryland 20899, USA}

\author{M.~B.~Stone}
\affiliation{Neutron Scattering Division, Oak Ridge National Laboratory, Oak Ridge, Tennessee 37831, USA}

\author{O.~Tchernyshyov}
\affiliation{Institute for Quantum Matter and Department of Physics and Astronomy, Johns Hopkins University, Baltimore, MD 21218, USA}

\author{S.~Nakatsuji}
\affiliation{Institute for Quantum Matter and Department of Physics and Astronomy, Johns Hopkins University, Baltimore, MD 21218, USA}
\affiliation{Institute for Solid State Physics (ISSP), University of Tokyo, Kashiwa, Chiba 277-8581, Japan}
\affiliation{CREST, Japan Science and Technology Agency (JST), 4-1-8 Honcho Kawaguchi, Saitama 332-0012, Japan}

\author{C.~Broholm}
\affiliation{Institute for Quantum Matter and Department of Physics and Astronomy, Johns Hopkins University, Baltimore, MD 21218, USA}
\affiliation{NIST Center for Neutron Research, National Institute of Standards and Technology, Gaithersburg, Maryland 20899, USA}
\affiliation{Department of Materials Science and Engineering, Johns Hopkins University, Baltimore, MD 21218, USA}

\date{\today}

\begin{abstract} 
We report the magnetic structure and spin excitations of \ce{Mn3Ge}, a breathing kagome antiferromagnet with transport anomalies attributed to Weyl nodes. Using polarized neutron diffraction, we show the magnetic order is a ${\bf k}={\bf 0}$ co-planar state belonging to a $\Gamma_9$ irreducible representation, which can be described as a perfect 120$\degree$ anti-chiral structure with a moment of 2.2(1)~$\mu_{B}$/\ce{Mn}, superimposed with weak collinear ferromagnetism. Inelastic neutron scattering shows three collective ${\bf Q}=0$ excitations at $\Delta_1$~=~2.9(6)~meV, $\Delta_2$~=~14.6(3), and $\Delta_3$~=~17.5(3)~meV. A field theory of $Q\approx 0$ spin waves in triangular antiferromagnets with a $120^\circ$ spin structure was used to classify these modes. The in-plane mode ($\alpha$) is gapless, $\Delta_1$ is the gap to a doublet of out-of-plane spin excitations ($\beta_x,\beta_y$), and $\Delta_2$, $\Delta_3$ result from hybridization of optical phonons with magnetic excitations. While a phenomenological spin Hamiltonian including exchange interactions, Dzyaloshinskii-Moriya interactions, and single ion crystal field terms can describe aspects of the Mn-based magnetism, spin wave damping ($\Gamma=25$(8)~meV) and the extended range of magnetic interactions indicate itinerant magnetism consistent with the transport anomalies.
\end{abstract}

%\pacs{75.30.Cr, 75.40.Cx, 75.50.Lk}% PACS, the Physics and Astronomy

\maketitle

%--------------------------------------------------------------------------------------------------------

\section{Introduction}
Non-collinear itinerant magnetism can give rise to anomalous electronic transport through its impacts on the Berry phase of itinerant electrons~\cite{Ye2018,Liu2018new,nakatsuji2015large,Kiyohara2016,nayak2016large}. The resulting coupling of electronic transport with magnetism presents important technological opportunities~\cite{doi:10.1063/1.5064697,doi:10.1063/1.5000815,Kimata2019,PhysRevB.95.075128,PhysRevB.99.184425,Liu2018} and new fundamental physics may arise from the impacts of Weyl points on magnetic interactions, phase transitions, and excitations. This field of topological magnetism is driven by the discovery of new materials with frustrated magnetic interactions that induce non-collinear magnetism, strong spin-orbit coupling, and a semi-metallic band structure with topologically protected Dirac or Weyl nodes near the chemical potential.

Fitting the bill, the hexagonal compounds \ce{Mn3X} (X= Sn/Ge) (space group $P6_{3}/mmc$, No.194) are semi-metallic antiferromagnets where Mn atoms form close-packed breathing kagome lattices with \ce{Sn/Ge} atoms at the center of Mn hexagons. Despite an apparent spontaneous magnetization of only $\approx$0.007~$\mu_B$/Mn, these antiferromagnets have large anomalous Hall (AHE) and Nernst effects (ANE) at room temperature with magnitudes comparable to strong ferromagnets~\cite{nakatsuji2015large,nayak2016large,Kiyohara2016,ikhlas2017large}.

Density Functional Theory indicates these transport anomalies arise from Weyl nodes near the chemical potential~\cite{kubler2014non,yang2017topological,kubler2018weyl}. Consistent with this, a chiral anomaly was discovered in magnetotransport measurements~\cite{nayak2016large,Kiyohara2016}. While their Weyl points are not pinned to the chemical potential, \ce{Mn3X} display many of the characteristics of a time-reversal-symmetry-breaking Weyl semi-metal ~\cite{kuroda2017evidence}.
Because their anomalous transport properties are not accompanied by a large ferromagnetic moment, they are formed from earth abundant elements, and they function at room temperature, $\rm Mn_3Ge$ and $\rm Mn_3Sn$ have serious application potentials~\cite{chappert2010emergence,shick2010spin,macdonald2011antiferromagnetic,park2011spin,gomonay2014spintronics,jungwirth2016antiferromagnetic}.

Previous diffraction studies notwithstanding~\cite{nagamiya1982triangular,tomiyoshi1983triangular,cable1993magnetic}, various magnetic structures are discussed in the recent literature. These are all based on an anti-chiral $120\degree$ structure but any in-plane easy axis cannot be determined through diffraction from a multi-domain structure~\cite{brown1990determination}. To overcome this problem, a recent polarized neutron diffraction study was performed on a field-cooled sample of \ce{Mn3Ge} and an in-plane easy axis along the [120] direction was found~\cite{PhysRevB.101.140411}. Here we specifically examine the in-plane magnetic spin order in the presence of a large magnetic field within the basal plane. Angle-resolved photoemission spectroscopy shows strong quasi-particle damping in \ce{Mn3Sn} indicative of electron-magnon interactions~\cite{kuroda2017evidence} but information about the nature of magnetic excitations is limited. Clearly, there is a need for precise knowledge of the magnetic structure, interactions, and excitations in \ce{Mn3X} to understand and exploit the anomalous transport.~\cite{yang2017topological,kondorskii1964degree,xiao2010berry}. 

In this manuscript, we determine the magnetic structure of \ce{Mn3Ge} through comprehensive polarized neutron diffraction experiments in zero and applied fields and we probe the low energy spin dynamics using time-of-flight neutron spectroscopy. As we searched for a minimal spin Hamiltonian to describe the collective magnetism in \ce{Mn3Ge}, we found it necessary to include exchange interactions well beyond nearest neighbors. The complexity of a lattice model with extended-range interactions and the lack of collective magnons beyond the $\Gamma$ point make a direct microscopic approach impractical. We thus took a different route to build a theoretical model by focusing on long-wavelength magnons that we describe through a field theory of spin waves in a continuous medium. This allows us to classify three modes at the $\Gamma$ point and quantitatively establish an effective low energy spin hamiltonian for \ce{Mn3Ge}. The analysis provides a template for understanding the long wavelength spin dynamics of triangular antiferromagnets.

The outline of the main manuscript is as follows: the experimental results are summarized in Section~\ref{sec.Result} including high-temperature specific heat measurements, polarized neutron diffraction experiments to determine the magnetic structure of \ce{Mn3Ge}, and time-of-flight neutron scattering experiments probing magnetic excitations and phonons. In Section~\ref{theorysection}, we describe a field theory of triangular antiferromagnetism, which is applied to \ce{Mn3Ge} and used to constrain an effective spin Hamiltonian in Section~\ref{HamiltonianMnGe}, before concluding in Section~\ref{conclusion}.

\section{Experimental Methods}\label{sec.ExpMet}

Single crystals of \ce{Mn3Ge} were obtained following a previously published protocol~\cite{Kiyohara2016}. Polycrystalline samples were prepared by arc melting the mixtures of manganese and germanium in a purified argon atmosphere. Excess manganese (at. 2.5$\%$) over the stoichiometric amount was added. The obtained polycrystalline materials were used for the crystal growth performed by the Bridgman–Stockbarger method. For this growth, the sample was heated up to 1050$^{\circ}$C and maintained at this temperature for 48 hours. Then, the sample was cooled to 740$^{\circ}$C at a rate of 5$^{\circ}$C/hour. The sample was annealed for 7 days at 740$^{\circ}$C and quenched in room temperature water to avoid precipitation of the low-temperature phase, which has the tetragonal Al$_3$Ti-type structure. The structure of \ce{Mn3Ge} was determined via single-crystal x-ray diffraction collected on a SuperNova diffractometer from Rigaku Oxford Diffraction and the data was refined using SHELXTL~\cite{sheldrick2015} at room temperature.

Specific heat data were acquired on a 6.90(1)~mg single-crystalline sample using the adiabatic method on a physical properties measurement system. The sample was fixed to the stage with Apiezon type H grease. The specific heat of the grease was separately measured and subtracted as a background to isolate the contribution from the sample.  

Polarized neutron diffraction experiments in zero and applied magnetic field were performed on the Triple-Axis Spectrometer (BT-7) at the NIST Center for Neutron Research (NCNR)~\cite{lynn2012double}. Neutrons with energy of 14.7~meV were selected for both incident and scattered beams. The single crystals were cooled to 10~K using a closed-cycle-refrigerator (CCR). Nuclear spin polarized $^3\ce{He}$ gas was used to polarize the incident neutron beam and analyze the polarization of the scattered beam~\cite{chen20073he}. Horizontal and vertical guide fields (HF and VF) were present throughout the beam path to allow measurements of neutron scattering cross-sections in two different polarization configurations. The spin-flip (SF) and non-spin-flip (NSF) scattering cross-sections were measured for incident neutron spins that are polarized parallel to the momentum transfer ${\bf Q}$ (HF) or perpendicular to the scattering plane (VF). The flipping ratio measured through nuclear Bragg diffraction from a pyrolytic graphite single crystal was $\sim$30-40 at the beginning of low field experiments and above 15 throughout the experiment.

Polarized neutron diffraction in a 2~T field perpendicular to the scattering plane was performed using an asymmetric split coil superconducting magnet resulting in a flipping ratio of $\sim$6-8 for the pyrolytic graphite sample. The polarization decay of the $^{3}$He cells was characterized as a function of time and this information was used to make time-dependent polarization corrections to the diffraction data~\cite{chen2011polarized}. 

In both zero and high-field diffraction experiments, two single crystals of \ce{Mn3Ge} were aligned on the same aluminum mount for simultaneous access to the ($h0\ell$) and the ($hh\ell$) reciprocal lattice planes. The (001) directions of the two crystals were intentionally offset from each other to distinguish their diffraction peaks. The Cooper-Nathans formalism was used to calculate the resolution function of BT-7~\cite{cooper1967resolution} and convert the integrated intensities of rocking scans to fully integrated Bragg intensities. 

Inelastic neutron scattering was measured using the fine resolution Fermi Chopper Spectrometer (SEQUOIA) at Oak Ridge National Laboratory (ORNL)~\cite{granroth2010sequoia}. Three single crystals with a total mass of $\sim$5 grams were co-aligned on an aluminum mount and installed in a CCR with a base temperature of 5~K. Inelastic neutron scattering spectra were acquired for a total sample rotation range of 194\degree~in 2\degree~increments for $E_i=22$~meV and $E_i=40$~meV, and a total sample rotation range of 214\degree~with 1\degree~steps for $E_i=300$~meV. Spectra with a total proton charge of 73.5~C, 147~C, and 322.5~C were collected for $E_i=22$~meV, $E_i=40$~meV, and $E_i=300$~meV respectively. We used the coarse Fermi chopper throughout rotating at $\nu~=~240$~Hz, $\nu~=~420$~Hz and $\nu~=~600$~Hz respectively. The inelastic neutron scattering data were reduced using Mantid~\cite{arnold2014mantid} and visualized using Horace~\cite{ewings2016horace}. The 4D resolution function was simulated via Monte Carlo ray tracing using McViNE~\cite{LIN201686,Lin2019}. 

\section{Results and Analysis}\label{sec.Result}
\begin{figure}[t]
    \includegraphics[width=\columnwidth]{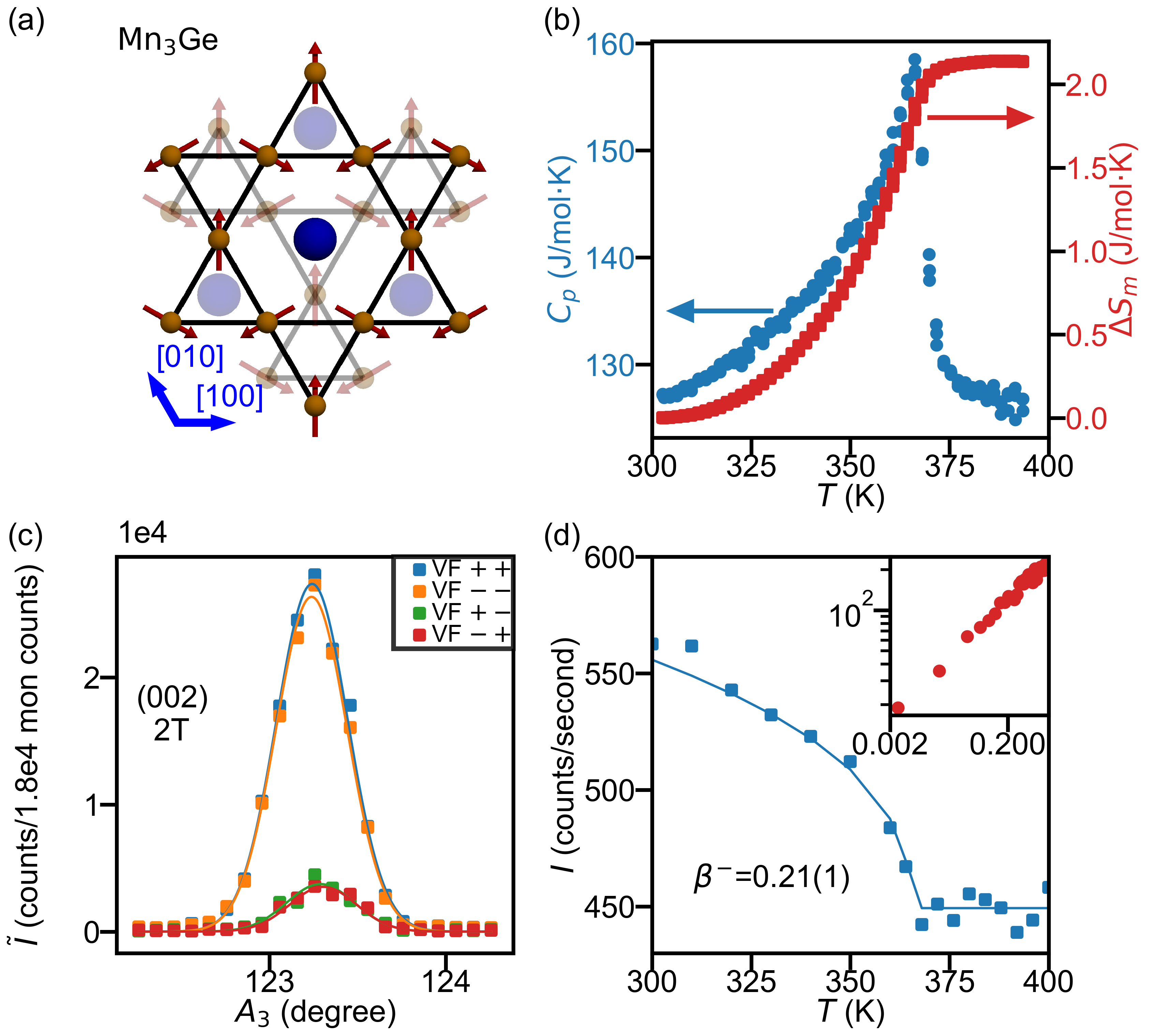}
    \centering
    \caption{(a) The refined anti-chiral magnetic structure of \ce{Mn3Ge} where the dark orange (blue) dots represent Mn (Ge) atoms. (b) The specific heat and the associated entropy released between 300~K and 400~K in \ce{Mn3Ge}. (c) Rocking scans through the $\mathbf{Q}~=~(002)$ Bragg peak for four different  polarized neutron scattering cross-sections measured in an applied field of 2~T. (d) is a neutron order parameter measurement collected on the $\mathbf{Q}~=~(001)$ Bragg peak of \ce{Mn3Ge} sensitive to magnetic diffraction through a multiple scattering process that involves the magnetic (302) reflection. Calculated and observed polarized beam Bragg diffraction cross-sections for \ce{Mn3Ge} are shown in (e) for zero field and in (f) for a 2~T field. For a closer inspection of these data see Table~\ref{tab:structurefactors}.  In (c), HF means the neutron polarization vector is parallel to the momentum transfer ${\bf Q}$, VF means the polarization vector is perpendicular to the scattering plane and $+/-$ refers to the spin states of the incident or scattered neutrons.}
    \label{fig: Fit}
\end{figure}
\subsection{Crystal structure of \ce{Mn3Ge}}\label{sec.ResultCrystal}
The $P6_{3}/mmc$ crystal structure of our \ce{Mn3Ge} single crystals was ascertained by single-crystal x-ray diffraction. The hexagonal structure is represented in Fig.~\ref{fig: Fit}(a) where Mn ions are dark orange and Ge ions are blue. The Mn ions form kagome layers that are stacked along the $c$ axis in an AB fashion with a layer spacing of $c/2$. The room temperature lattice parameters are $a=b=5.3315(5)$~\AA\ and $c=4.3055(3)$~\AA\ and the ideal structure contains 6 symmetrically equivalent Mn ions. 
Mn is one of the few atoms with a negative neutron scattering length so the contrast $|(f_{\ce{Ge}}-f_{\ce{Mn}})/(f_{\ce{Ge}}+f_{\ce{Mn}})|$ between Ge and Mn is much larger for neutrons (2.67) than for non-resonant x-ray photons (0.12). We thus determined the stoichiometry of our Mn$_{3+x}$Ge$_{1-x}$ crystal based on neutron diffraction, which refined to \ce{Mn_{3.07}Ge_{0.93}}. This becomes important in the magnetic structure refinement as magnetic and nuclear scattering both contribute to non-spin-flip vertical field diffraction (see Eqs.~\ref{nsfp} and \ref{nsfm} in Appendix A). 

\subsection{Magnetic structure of \ce{Mn3Ge}}\label{sec.Result1}
The temperature dependence of the specific heat (Fig.~\ref{fig: Fit}(b)) indicates a second-order phase transition at $T_{N}=365$~K with a change in entropy of $\Delta S_{m}=2.1$~J/mol/K per formula unit that only amounts to 4.7(1)\% of the total entropy of three spin-5/2 manganese atoms ($3R\ln 6$). The sharp nature of the anomaly and the fact that resolution limited magnetic Bragg peaks appear at the same critical temperature shows that the specific heat peak is associated with the majority phase. For comparison, the critical temperature of the tetragonal minority phase is 800~K~\cite{doi:10.1143/JPSJ.16.1995}. The reduced $\Delta S_m$ is consistent with the reduced ordered moment of Mn$_3$Ge (see below) as both can result from a partial gapping of the Fermi surface in an itinerant description of the magnetism. Alternatively, the reduction in $\Delta S_m$ could result from persistent short-range spin correlations above $T_N$ that can arise from competing interactions. 

Consistent with previous findings~\cite{nagamiya1982triangular,tomiyoshi1983triangular}, the onset of spin-flip (SF) diffraction at structural Bragg peaks for  $T<T_{N}$ indicates ${\bf k}={\bf 0}$ antiferromagnetic (AFM) ordering. We obtain the order parameter critical exponent $\beta=0.21(1)$ from the temperature dependence of the Bragg scattering intensity in the critical regime (Fig.~\ref{fig: Fit}(d)). Similar values have been measured in other non-collinear antiferromagnets with triangular lattices such as \ce{CsMnBr3} ($\beta=0.21(2)$)~\cite{PhysRevB.39.586}, \ce{VCl2} ($\beta=0.20(2)$)~\cite{doi:10.1143/JPSJ.56.4027}, \ce{K2CuF4} ($\beta=0.22$)~\cite{doi:10.1143/JPSJ.35.1328} and \ce{Mn(HCOO)$_2\cdot 2$H$_2$O} ($\beta=0.23(1)$)~\cite{PhysRev.188.1037}. The critical exponent is close to the $U(1)\times Z_2$ universality class with $\beta=$0.25-0.28~\cite{doi:10.1063/1.340905} which is, however a 2D model while \ce{Mn3Ge} is clearly a 3D system. Monte Carlo simulation of a 3D stacked triangular antiferromagnet with 3D Heisenberg interactions found a critical exponent $\beta=0.221(9)$, which is within error bars of our experimental value~\cite{MC_sim}.

\begin{figure}[t]
    \includegraphics[width=1\columnwidth]{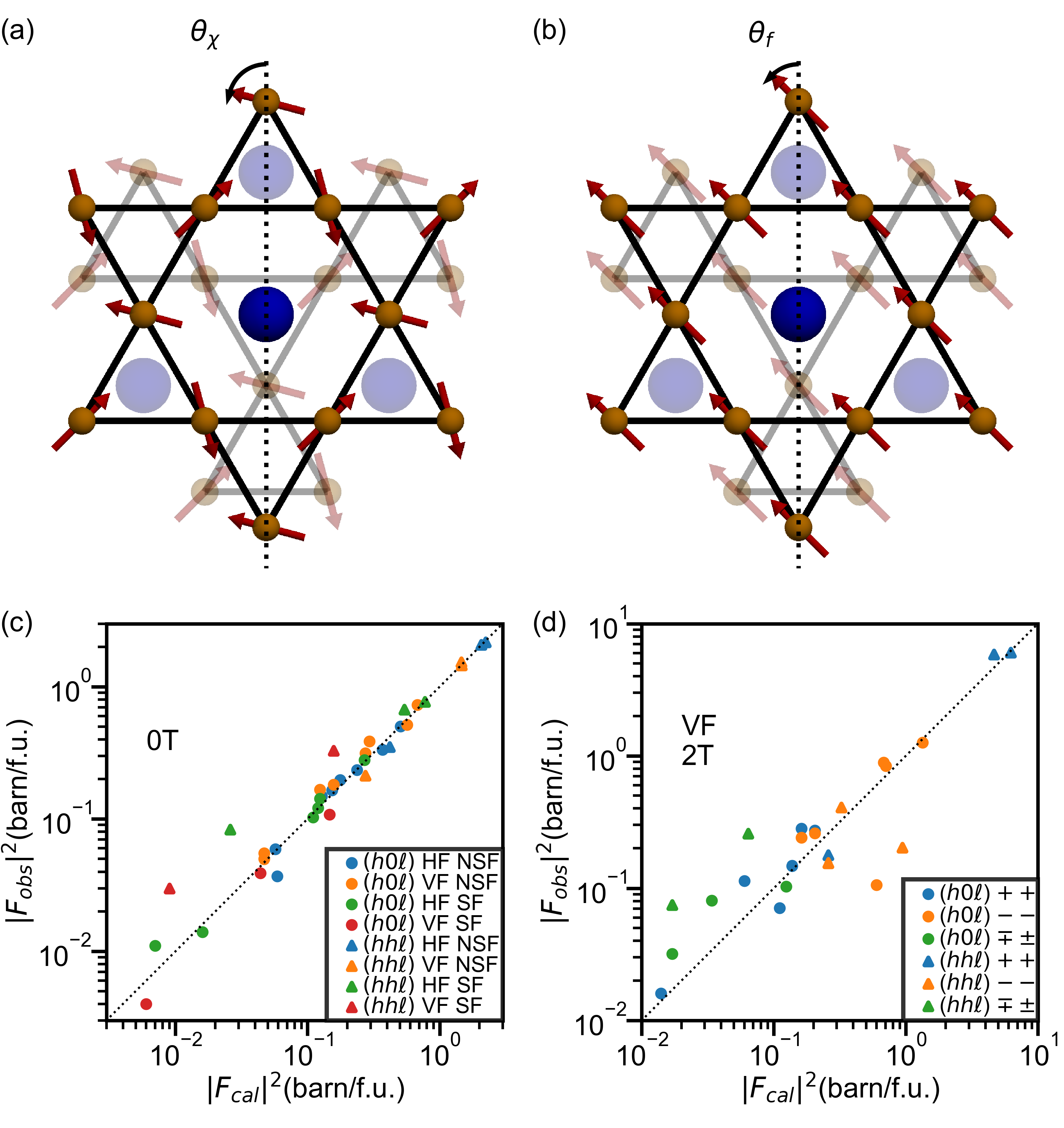}
    \centering
    \caption{The magnetic structure of \ce{Mn3Ge}, which is associated with the $\Gamma_{9}$ irreducible representation and can be decomposed into (a) an anti-chiral component and (b) a ferromagnetic component. The in-plane orientation of each component of the spin structure is defined by a global rotation angle for all spin, which we define as $\theta_{\chi}$ and $\theta_f$ for the anti-chiral and ferromagnetic components respectively, and as indicated on the figure. The dark orange (blue) dots represent the Mn (Ge) ions. Calculated and observed polarized beam Bragg diffraction cross-sections for \ce{Mn3Ge} are shown in (c) for zero field and in (d) for a 2~T field. For a closer inspection of these data see Table~\ref{tab:structurefactors}. HF means the neutron polarization vector is parallel to the momentum transfer ${\bf Q}$, while VF means the polarization vector is perpendicular to the scattering plane, while $+/-$ refers to the spin states of the incident or scattered neutrons.}
    \label{fig: IRs}
\end{figure}

\begin{figure*}
    \includegraphics[width=7in]{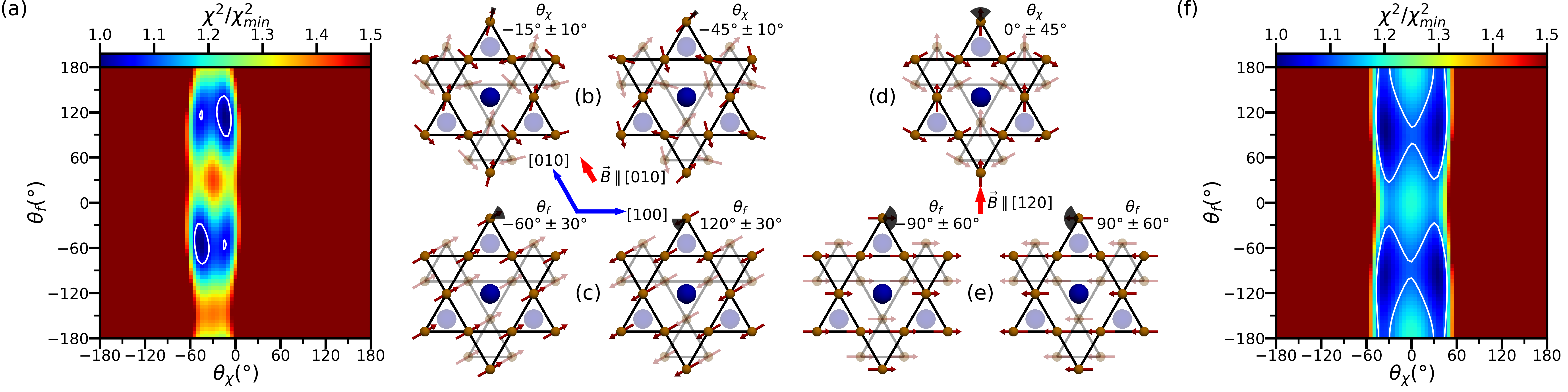}

    \caption{Angular parameters associated with the anti-chiral ($\theta_{\chi}$) and the ferromagnetic  ($\theta_{f}$) components of the magnetic order in Mn$_3$Ge in a 2~T field (a,b,c) {\bf B}$\parallel$[010] and (d,e,f) {\bf B}$\parallel$[120] as determined by polarized and unpolarized neutron diffraction. (a) and (f) show the $\chi^2$ goodness of fit for the two field directions as a function of the angular parameters $\theta_\chi$ and $\theta_f$ that define the magnetic structures as depicted in Fig.~\ref{fig: IRs}. The white lines in (a) and (f) indicate the experimental constraints on $\theta_{\chi}$ and $\theta_{f}$. (b) and (d) represent the anti-chiral components of the magnetic structure refined for each of the two field directions, while (c) and (e) are the corresponding refined ferromagnetic components.}
    \label{fig: fielddata}
\end{figure*} 

The second-order nature of the phase transition ensures the magnetic structure forms an irreducible representation (IR) of the space group. Of the ten IRs listed in Table~\ref{tab:chararctertable}, only $\Gamma_9$ is consistent with an in-plane ferromagnetic moment~\cite{nayak2016large,Kiyohara2016} and our observation of magnetic diffraction at ${\bf Q}$ = (002) (Fig.~\ref{fig: Fit}(c)). The $\Gamma_{9}$ magnetic structures can be described as a superposition of a perfect anti-chiral triangular structure that has no net magnetization with a ferromagnet polarized within the basal plane. This decomposition is depicted in Fig.~\ref{fig: IRs}. $\Gamma_9$ accommodates separate continuous rotations of the anti-chiral and the ferromagnetic structures within the basal plane. The structure can thus be parametrized by $(M_{\chi}, M_{f}; \theta_{\chi}, \theta_{f})$. $M_{\chi}>0$ and $M_{f}>0$ are the moment sizes of the anti-chiral component and the net ferromagnetic moment per Mn atom respectively. As defined in Fig.~\ref{fig: IRs}(a), $\theta_{\chi}$  describes the counter clockwise in-plane rotation of every anti-chiral spin component relative to the perfect anti-chiral version of the structure shown in Fig.~\ref{fig: Fit}(a) while $\theta_{f}$ describes the counter-clockwise in-plane rotation of the ferromagnetic (FM) component.  Assuming equal volume fractions of the rotational domains generated by the original 6-fold symmetry and 2-fold symmetry about [001] and $\langle110\rangle$ respectively of the ordered state, the Bragg diffraction cross-section depends only on $M_{\chi}$, $M_{f}$ and $\theta_{\chi}+\theta_{f}$. Thus, magnetic structures in the $U(1)$ manifold $(M_{\chi},M_{f};\theta_{\chi}+\theta,\theta_{f}-\theta)$ are indistinguishable through diffraction from a multi-domain state. The actual zero-field domain distribution of our \ce{Mn3Ge} sample could not be refined based on the available data. However, the magnetic domains are energetically equivalent and a multi-domain is favored by dipole-dipole interactions. A significant deviation from an equiprobable distribution of the magnetic domains is thus unlikely.

To determine $\theta_{\chi}$ and $\theta_{f}$~\cite{brown1990determination}, we performed polarized neutron diffraction experiments with a 2~T magnetic field applied along the [010] and the [$1\bar{1}$0] directions respectively to shift the domain population. The magnetic field was perpendicular to the $(h0\ell)$ and $(hh\ell)$ scattering planes respectively. Four polarized cross-sections were measured for each Bragg peak corresponding to the incident and scattered neutrons polarized either parallel ($+$) or anti-parallel ($-$) to the applied magnetic field. Excluding Bragg peaks with contributions from multiple scattering (such as (001) and (101)), we then combined in-field polarized diffraction data with the VF and HF cross-sections of the multi-domain state to determine $(M_{\chi},M_{f};\theta_{\chi},\theta_{f})$.

The comparisons between the observed and the best fit structure factors in 0~T and 2~T are shown in Fig.~\ref{fig: IRs}(c) and Fig.~\ref{fig: IRs}(d) respectively and in Table~\ref{tab:structurefactors}. We note that for a few Bragg peaks in the $(hhl)$ plane there are relatively large discrepancies between measurements and calculations, which we associate with multiple scattering as described in Appendix A. From the multi-domain data we obtain $M_{\chi}=2.2(1)\mu_{B}$ per Mn and $M_{f}$=0.2(1)$\mu_{B}$ per Mn$_3$Ge for the ferromagnetic component. $M_{\chi}$ is consistent with previous neutron diffraction results while $M_{f}$ is almost two orders of magnitude greater than the value obtained from magnetization measurements~\cite{nayak2016large,Kiyohara2016}. Fig.~\ref{fig: fielddata} shows the $\chi^2$ goodness of fit versus the angular parameters for two field orientations. A macroscopic sample will in general contain symmetry restoring domains that map onto each other through the action of paramagnetic space group symmetry operations that remain in the presence of any applied magnetic field. For {\bf B}$\parallel$[120] there are four domains indexed as 11, 12, 21, and 22, which are related by $\theta_{\chi,f}^{(2)}$~=~$-\theta_{\chi,f}^{(1)}$. The corresponding expressions for {\bf B}$\parallel$[010] are $\theta_{\chi,f}^{(2)}= -60\degree-\theta_{\chi,f}^{(1)}$. In the refinement of the high field polarized diffraction data, we superimposed contributions from each of these domains. For both field directions, the anti-chiral spin state consists of domains of the type shown in Fig.~1(a). The three Mn spins on the vertices of each triangle are generally oriented at $120\degree$ to each other with one spin bisecting the triangle. Even in the presence of a field along [010] a single domain $\theta_{\chi}=90\degree$ state is inconsistent with the data. This result is consistent with a recent neutron polarisation experiment on a field-cooled sample of \ce{Mn3Ge}~\cite{PhysRevB.101.140411}. 

\begin{figure*}
    \includegraphics[width=7in]{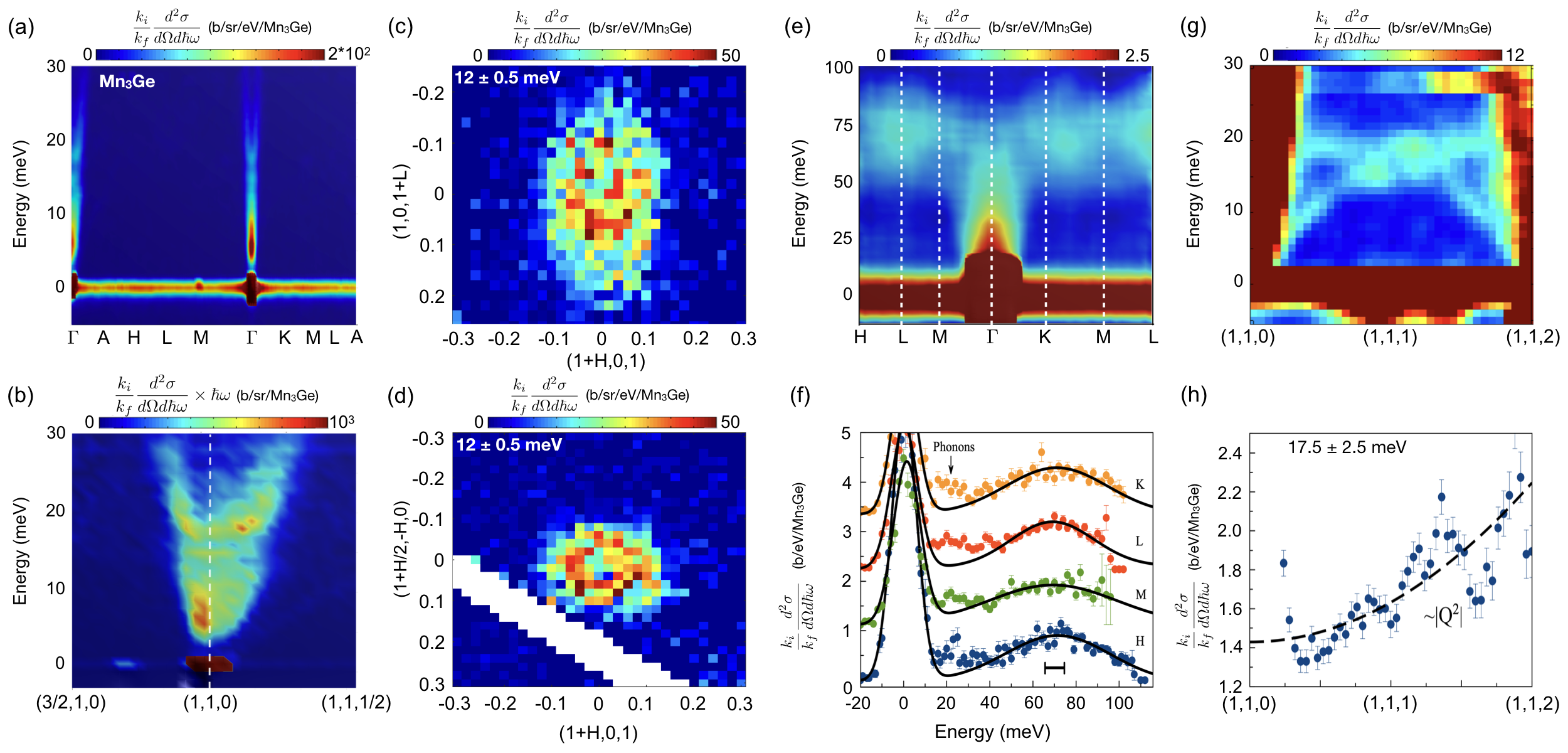}

    \caption{Time-of-flight neutron scattering data for \ce{Mn3Ge} collected for $E_i=40$~meV (a,b,c,d,f,h) and for $E_i=300$~meV (e,f). (a) inelastic neutron scattering from \ce{Mn3Ge} acquired for $E_i=40$~meV for wave vector transfer along high symmetry trajectories of the hexagonal Brillouin zone. (b) The energy transfer weighted neutron scattering cross-section at low energies and for momentum transfer near the zone center. Both (a) and (b) show data averaged over $\pm$0.05$~\angstrom^{-1}$ in the perpendicular $\mathbf{Q}$ directions. (c) Shows a constant energy slice within the $(h0\ell)$ plane centered at $\mathbf{Q}~=~(101)$, and (d) within the $(hk0)$ plane. Data from (c) and (d) were averaged over $\pm$0.07$~\angstrom^{-1}$ in the perpendicular momentum directions. High energy inelastic neutron scattering acquired for $E_i=300$~meV and wave vector transfer along high symmetry trajectories of the hexagonal Brillouin zone. (f) Energy cuts at 4 different symmetry points are plotted in (f) where the horizontal line represents the FWHM energy resolution. Panel (g) shows the $(00\ell)$ dependence of inelastic neutron scattering between two magnetic zone centers, which reveals acoustic phonon dispersion. The momentum dependence of the neutron scattering cross-section from phonons is shown in (h), where the black dashed line is proportional to $|\mathbf{Q}|^2$. The error bars in (h) and in all other figures of the paper correspond to 1 standard deviation. The data in panels (e,f,g,h) were averaged over $\pm$0.1~$\angstrom^{-1}$ in perpendicular $\mathbf{Q}$ directions.}
    \label{fig: lowEspectra}
\end{figure*} 

The ferromagnetic component refines to be perpendicular to the applied field for both field directions. This transverse nature of the uniform magnetization inferred from the polarized diffraction data and reported in Fig.~\ref{fig: fielddata}(c,e) is surprising. Fig.~\ref{fig: Fit}(c) shows the relevant rocking scans at $\mathbf{Q}$~=~(002) for 4 different polarization channels after correction for finite beam polarization effects. Bragg diffraction in the two spin-flip channels is clearly present at (002). Possible extraneous sources of spin-flip scattering at (002) are discounted based on the following considerations. 

(i) Multiple scattering at (002) is forbidden for 14.7~meV neutrons when the incident and scattered momenta lie in the $(h0\ell)$ plane. To confirm this we rotated the sample around (002) in a $\chi$ scan and found the spin-flip Bragg scattering to be independent of {$0\degree<\chi<14\degree$}, which is inconsistent with a multiple scattering process (Fig.~\ref{fig:chi_square}(d)). 

(ii) Depolarization effects caused by the guide field configuration surrounding the 2~T superconducting magnet were checked by measuring the flipping ratio for a PG sample placed at the sample location during the experiment. The resulting flipping ratio was found to be ~8 at the beginning of the polarized experiments, which is consistent with the value used in the polarization corrections. 

(iii) The effects of sample depolarization were investigated by performing a refinement of the polarized diffraction data with a sample depolarization factor chosen such that the intrinsic spin-flip scattering at (002) is zero. This refinement gives the best fit with a zero ferromagnetic moment, but with a $\chi^{2}$ value that is $8\%$ larger than for a fit without sample depolarization. This indicates that neutron spin-flip scattering at (002) is an intrinsic property of our sample.

Intrinsic magnetic Bragg diffraction at $\mathbf{Q}$~=~(002) arises from a ferromagnetic moment within the basal plane of Mn$_3$Ge. As apparent in the polarized neutron scattering cross-sections listed in the SI, a difference between the $--$ and $++$ cross-sections is directly proportional to the squared component of magnetization along the applied field. In Fig.~\ref{fig: Fit}(c), the rocking scans for the $--$ and $++$ polarization configurations are indistinguishable. This is consistent with the magnetization of just 0.007~$\mu_B$/Mn along the applied field determined by SQUID magnetometry~\cite{Kiyohara2016,nayak2016large}. spin-flip Bragg scattering at (002) on the other hand, probes uncompensated magnetization perpendicular to the applied field within the coherence volume of each basal plane. The sample averaged perpendicular uncompensated magnetization of 0.2(1)$\mu_B$/Mn that we detect is not inconsistent with the much smaller longitudinal magnetization component seen in SQUID magnetometry. It could result from a minority phase or an orbital moment as discussed in Appendix B.

\subsection{Magnons and Phonons in \ce{Mn3Ge}}\label{sec:SpinPho}
Fig.~\ref{fig: lowEspectra}(a) shows the momentum ($\bf Q$) and energy dependence of the inelastic neutron scattering cross-section for $\bf Q$ traversing high symmetry trajectories through the Brillouin zone and for energy transfer up to 30~meV. The data show that most of the long-wavelength low energy magnetic spectral weight is associated with linearly dispersive excitations emanating from each magnetic zone center $\Gamma$.

Fig.~\ref{fig: lowEspectra}(b) shows the energy transfer weighted scattering cross-section near the magnetic zone center. Three distinct modes are observed at $\Delta_1$~=~2.9(6)~meV, $\Delta_2$~=~14.6(3), and $\Delta_3$~=~17.5(3)~meV, but a single linearly dispersive branch is observed above $\approx$20~meV. 12~meV constant energy maps in the $(h0\ell)$ plane (Fig.~\ref{fig: lowEspectra}(c)) and the $(hk0)$ plane (Fig.~\ref{fig: lowEspectra}(d)) near ($\bf Q$)~=~(101) show a single well-defined ellipsoid in momentum space. The eccentricity of the ellipsoid in Fig.~\ref{fig: lowEspectra}(c) indicates 1.8(1) times faster magnon velocity within than perpendicular to the basal plane while Fig.~\ref{fig: lowEspectra}(d) indicates isotropic dispersion within the basal plane at low energies. From the principal axis lengths, we obtain $c_{\beta}^L$~=~92(8)~meV$\cdot$\AA~~along the out-of-plane $(00\ell)$ direction, while we get $c_{\alpha}^H$~=~170(12)~meV$\cdot$\AA~for the in-plane $(h00)$ direction. Here the $(00\ell)$ and $(h00)$ directions respectively correspond to the $\Gamma-A$ and $\Gamma-M$ directions within the first Brillouin zone of the hexagonal structure.

Fig.~\ref{fig: lowEspectra}(e) shows the continuation of the spin excitations to high energy where a flat band of scattering is centered around 75~meV and magnetic spectral weight is observed up to at least 100~meV. The corresponding high energy spectra for select symmetry points within the Brillouin zone are in Fig.~\ref{fig: lowEspectra}(f). A broad maximum centered between 70~meV and 75~meV is observed at all high symmetry points. By fitting the energy cut data to a Lorentzian, an intrinsic half-width at half maximum relaxation rate of 25(8)~meV was obtained. For comparison the full width at half maximum (FWHM) energy resolution (horizontal black bar in Fig.~\ref{fig: lowEspectra}(f)) is 8.8~meV. 

Parts of the phonon dispersion relations for \ce{Mn3Ge} were also characterized by inelastic neutron scattering. Linearly dispersive acoustic phonon branches originating from each $\Gamma$ points are apparent in the $(00\ell)$ dependence of the neutron scattering spectrum between two magnetic zone centers (Fig.~\ref{fig: lowEspectra}(g)). The top of the acoustic phonon band is near 15~meV, where optical phonons are also detected. The vibrational nature of these excitations is indicated by the $|\mathbf{Q}|^2$ dependence of their intensity (Fig.~\ref{fig: lowEspectra}(h)), which is consistent with the one-phonon scattering cross-section and contrary to a magnetic cross-section that decreases with $|\mathbf{Q}|$ due to the magnetic form factor. 

\section{Theory}\label{theorysection}
\subsection{Origin of non-collinear magnetism in Mn$_3$Ge}\label{sec:NonCol}
\noindent

\begin{figure}[t]
    \includegraphics[width=\columnwidth]{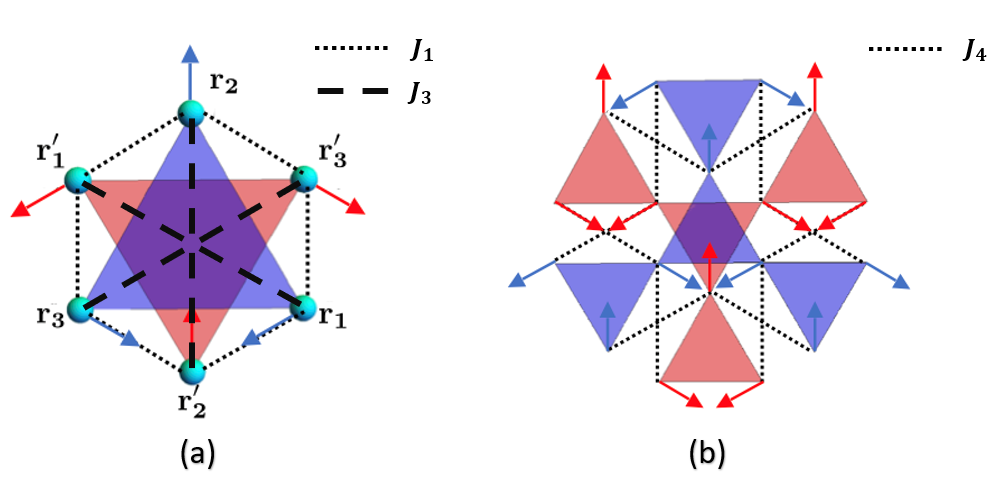}
    \centering
    \caption{Inter-layer exchange interactions (dashed and dotted lines) between blue spins in the blue bottom layer and red spins in the red top layer. Spins are oriented as in the long-range ordered structure of Mn$_3$Ge. (a) $J_1$, which is weak and antiferromagnetic for Mn$_3$Ge is marked with dotted lines, and $J_3$, which is ferromagnetic is marked with dashed lines. We drop $J_3$ in our model as it produces a similar effect on the spin wave dispersion as $J_1$.  (b) $J_4$, which is strong and ferromagnetic for Mn$_3$Ge is indicated by dashed lines.}
    \label{fig:interplane-interaction}
\end{figure}

Insights into the magnetism of \ce{Mn3Ge} can be obtained by considering the following spin Hamiltonian, which we shall denote as the JDK model:
\begin{eqnarray}\label{eq:HCEF}
\mathcal{H}_{JDK}=&& \sum_{<i,j>} J_{ij}~ \mathbf{S}_i\cdot \mathbf{S}_j + \sum_{<i,j>} \mathbf{D}_{ij} \cdot (\mathbf{S}_i \times \mathbf{S}_j)
\nonumber\\
&-& \sum_{i} K(\uvec{n}_i \cdot \mathbf{S}_i)^2.
\end{eqnarray}
Mn ions in \ce{Mn3Ge} form a breathing kagome plane, however, we shall assume exchange interactions are unaffected by the small breathing amplitude ($\sim$ 0.02~\AA). The first term describes Heisenberg exchange interactions between Mn spins. For \ce{Mn3Ge}, the first nearest neighbor is out-of-plane (Fig.~\ref{fig:interplane-interaction}) and the second nearest-neighbor is in-plane. We shall denote the corresponding exchange interaction constant by $J_1$ and $J_2$ respectively. We note that the introduction of longer range inter-plane interactions is necessary to account for spin dynamics in \ce{Mn3Ge}. $J_3$ describes inter-plane interactions between Mn ions such as $\mathbf{r_1}$ and $\mathbf{r_1'}$ (dashed lines in Fig.~\ref{fig:interplane-interaction}(a)). The effects of $J_3$ on the spin wave dispersions of \ce{Mn3Ge} are similar to $J_1$ so we ignored $J_3$ in our analysis. The $J_4$ inter-layer interactions are indicated by black dotted lines in Fig.~\ref{fig:interplane-interaction}(b). The second term in Eq.~\ref{eq:HCEF} describes Dzyaloshinskii-Moriya (DM) interactions between the in-plane nearest-neighbor Mn spins. Symmetries of the \ce{Mn3Ge} lattice restrict the DM vectors ($\mathbf{D}_{ij}$) to point along the $c$ axis ($\mathbf{D}_{ij}=D\uvec{z}$). The last term describes single-ion anisotropy arising from the crystal field and spin-orbit coupling. The unit vector $\uvec{n}_i$ is parallel to the straight line that connects the centers of the two triangles that share a vertex at the spin site. As we shall see, $\mathbf{D}_{ij}$ and $K>0$ make the basal plane a macroscopic easy plane and open a gap in the out of plane spin wave excitation spectrum. 

With nearest-neighbor interactions only, the 2D kagome antiferromagnet is highly frustrated~\cite{chalker1992hidden}: There is no phase transition for Ising spins, the classical XY-model is critical, the classical Heisenberg model has a manifold of low energy states~\cite{ISI:A1992HL99300085}, and the spin-1/2 quantum Heisenberg model has several nearly degenerate valence bond solid and spin liquid states~\cite{RevModPhys.89.025003}. The fact that Mn$_3$Ge orders magnetically at high temperatures and supports magnons propagating in all directions, indicates interactions extend well beyond the nearest neighbors, as expected for an itinerant magnet. The DM interactions promote a 120\degree~structure where spins are constrained to basal-planes and only a macroscopic $U(1)$ degeneracy is preserved~\cite{elhajal2002symmetry}. The anti-chiral spin structure implies all the ${\bf D}_{ij}$ vectors point into the page when we adopt a clockwise indexing convention for the triangles that make up the kagome lattice. 

In previous work~\cite{cable1993magnetic,park2018magnetic}, $U(1)$ symmetry breaking was associated with a 6$^{th}$ order single-ion term because the 2$^{nd}$ and 4$^{th}$ order terms cannot break the $U(1)$ degeneracy of a perfect 120\degree~spin structure. However, the $\Gamma_9$ IR includes the possibility of in-plane canting and different moment sizes for the different Mn atoms. When the magnetic structure deviates from a perfect 120\degree~ spin structure, the second-order CEF term breaks the $U(1)$ degeneracy~\cite{liu2017anomalous}.

\subsection{Field Theory for \ce{Mn3Ge}}\label{sec:Field-theory-triang-AFM}

\subsubsection{Preliminary remarks}\label{sec:Field-theory-triang-AFM1}

Given the absence of well-defined magnons beyond the long-wavelength limit in this long-range ordered itinerant magnet, we cannot expect to determine a detailed microscopic spin hamiltonian for \ce{Mn3Ge}. Instead we formulate a continuum theory of spin waves that focuses on the universal long-wavelength features of the excitation spectrum that are accessible in our neutron scattering data~\cite{DasguptaMnGe}. The low energy collective excitations in the long-range ordered state are interpreted as linearly dispersive Goldstone modes arising from spontaneous spin rotation symmetry breaking. 

The general setting is an antiferromagnet with Heisenberg exchange interactions on a two-dimensional lattice with a triangle as a building block. We assume that classical ground states have a magnetic unit cell with three coplanar spins $\mathbf S_1$, $\mathbf S_2$, and $\mathbf S_3$ such that 
\begin{equation}
\mathbf S_1 + \mathbf S_2 + \mathbf S_3 = 0.
\label{eq:vacuum}
\end{equation}

\begin{figure}[t]
    \includegraphics[width=\columnwidth]{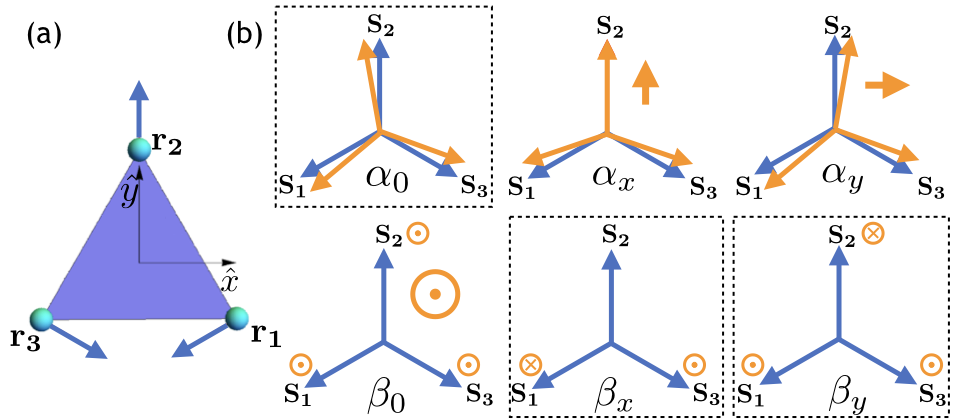}
    \centering
    \caption{(a) A single plaquette of the Mn sites in Mn$_3$Ge showing the ground state spin structure with blue arrows. The spins carry the same labels as the site i.e spin $S_i$ is at site $r_i$. (b) The normal modes for the spin structure with the in plane $\alpha$ modes, the blue arrows indicate the ground state, while the orange arrows indicate the distorted state on the top. The out of plane $\bm{\beta}$ modes are shown at the bottom. The three hard modes carry net spin indicated by the thick orange arrow beside $(\alpha_x,\alpha_y)$ and the out of plane arrow beside the $\beta_0$ mode. The three soft modes are surrounded by a dashed line.}
    \label{fig:lattice_modes}
\end{figure}

\subsubsection{Soft modes}\label{sec.Soft-modes}
\noindent
Figure \ref{fig:lattice_modes} shows the normal modes representing infinitesimal spin rotations in a single triangular unit cell (or in a spin wave with $\mathbf k=0$). There are three soft modes $\alpha_0$, $\beta_x$, and $\beta_y$ that represent global rotations of the three spins and thus preserve the ground-state condition (\ref{eq:vacuum}). The remaining 3 modes $\beta_0$, $\alpha_x$, and $\alpha_y$ are hard. From the perspective of the $D_3$ point group symmetry, $\alpha_0$ and $\beta_0$ transform under the trivial singlet representation; $(\alpha_x,\alpha_y)$ and $(\beta_x,\beta_y)$ are irreducible doublets transforming as a vector such as $\mathbf k = (k_x, k_y)$. 

\subsubsection{Field theory for the soft modes}
\noindent
The singlet mode $\alpha_0$ has simple dynamics. Its Lagrangian density consists of kinetic energy with mass density $\rho_\alpha$ and a potential energy quadratic in the gradients of $\alpha_0$:  
\begin{equation}\label{eq:L-alpha0}
\mathcal L = 
    \frac{\rho_\alpha}{2}  \dot{\alpha}_0^2 
    - \frac{\kappa}{2}  
        \partial_i \alpha_0 \,
        \partial_i \alpha_0.
\end{equation}
The summation is assumed to be performed over repeated Cartesian indices $i = x,y$. As often happens in highly symmetric solids, the effective Lagrangian (Eq.~\ref{eq:L-alpha0}) obeys not just the discrete symmetries of the point group $D_3$ but also the full rotational symmetry SO(2). Spin waves have a linear dispersion $\omega = c k$ with the speed $c = \sqrt{\kappa/\rho_\alpha}$. 

The continuum theory for the doublet is more involved as the doublet field $\bm \beta$ itself transforms under rotations. The Lagrangian of this field has the following form: 
\begin{equation}\label{eq:L-beta-explicit}
\mathcal L = 
    \frac{\rho_\beta}{2}  \dot{\beta}_i^{2} 
    - \frac{\lambda}{2} \, 
    \partial_i \beta_i \, \partial_j \beta_j
    - \frac{\mu + \tilde{\mu}}{2} 
        \partial_i \beta_j \partial_i \beta_j
    - \frac{\mu - \tilde{\mu}}{2} 
        \partial_i \beta_j \partial_j \beta_i.
\end{equation}
This structure is highly reminiscent of a continuum theory of elasticity, which we further elucidate in ref.~\cite{DasguptaMnGe}.

\subsubsection{Velocities of the soft modes for \ce{Mn3Ge}}
The 2D theory for the $\alpha_0$ and $\bm{\beta}$ modes can be extended with some modifications to the 3D structure of \ce{Mn3Ge}. The compound is a layered AB stacked Kagome system, within each layer the lattice parameter is given by the constant $a$, and the separation in the $z$-direction between two kagome planes is $l=c/2$. In each layer, the ground state has spins confined to the corresponding plane, and the three spins forming each AFM triangular plaquette are not rotated by exactly $2\pi/3$ with respect to each other such that a small FM moment appears. This slight deviation of the spins from a perfect $2\pi/3$ anti-chiral state, which is due to single-ion anisotropy, was not considered in the field theory.

An effective description of the system requires two sets of modes: $(\alpha_{0},\bm{\alpha},\beta_{0},\bm{\beta})$ for the A layer and $(\alpha'_{0},\bm{\alpha}',\beta'_{0},\bm{\beta}')$ for the B layer. The theory is better expressed in terms of symmetric and antisymmetric combinations of the two sets, $\zeta^{s} = \frac{\zeta + \zeta'}{\sqrt{2}}$ and $\zeta^{a} = \frac{\zeta - \zeta'}{\sqrt{2}}$, where $\zeta$ stands for any of the $\alpha$ or $\beta$ fields. The primary unit is the nuclear cell motif of \ce{Mn3Ge} that consists of an ``up triangle" in the lower (blue) layer and a ``down triangle" in the upper (red) layer (Fig.~\ref{fig:interplane-interaction}(a)).

The spins in each layer interact amongst themselves through the antiferromagnetic Heisenberg exchange interaction $J_2$. Inter-layer interactions can be FM or AFM and those that appear relevant in Mn$_3$Ge are shown in Fig.~\ref{fig:interplane-interaction}. The detailed description of the theory is presented in ref.~\cite{DasguptaMnGe}. Here we collect the results most relevant to the experiment, namely expressions for the in-plane and the out of plane magnon velocities and the associated energy gaps in terms of Hamiltonian parameters.\\
The in-plane velocities for the soft symmetric $\alpha$ mode and the two $\bm{\beta}$ modes respectively are:
\begin{eqnarray}\label{eq.velocities_total}
c_{\alpha}^{s}(\hat{k}_a) &=& \sqrt{\frac{1}{\rho_{\alpha^{s}}}\left(\frac{J_2}{8} + \frac{J_4}{3} + \frac{J_1}{24}\right)}aS\\ \nonumber
c_{||}^{s} (\hat{k}_a) &=& \sqrt{\frac{1}{4\rho_{\beta^{s}}}\left(J_2 + \frac{5J_4}{6} - \frac{J_1}{3} + \frac{3 J_1 J_4}{2(J_1+2J_4)}\right)}aS \\ \nonumber
c_{\perp}^{s} (\hat{k}_a) &=& \sqrt{\frac{3J_4}{8\rho_{\beta^{s}}}\left( 1  + \frac{J_1}{(J_1+2J_4)}\right)}aS. 
\end{eqnarray}
The inertias $\rho_{\alpha^{s}}$ and $\rho_{\beta^{s}}$ are given by 
\begin{equation}
\rho_{\beta^{s}} =  \frac{1}{3(J_2 + J_1)} = 2\rho_{\alpha^{s}}.
\end{equation}
Note that with just a $J_1$ out of plane interaction ($J_4 = 0$) the mode that was dispersionless under $J_2$ remains flat to linear order, but develops weak  dispersion along the $(hh0)$ direction at the quadratic level while remaining dispersionless along the $(h00)$ direction. $J_4$ on the other hand produces an isotropic propagating mode out of the flat kagome mode associated with $J_2$. 

The out of plane dispersion is set by the primed fields since according to our schema the unprimed fields are at $z = 0$. For the $\alpha'_0$ mode, the dispersion is given by $\rho_{\alpha}\omega_{\alpha}^{2} = ( \frac{J_4}{2} + \frac{J_1}{4}) (k_z l)^{2} $. For the $\beta'_{x,y}$ modes the c-dispersion is  $\rho_{\beta}\omega_{\beta}^{2} = ( \frac{J_4}{2} + \frac{J_1}{4}) (k_z l)^{2} $. These give the out of plane velocities as:
\begin{eqnarray}\label{eq.outofplanevelocities}
    c_{\alpha}^{s}(\hat{k}_c) = \sqrt{\frac{2J_4 + J_1}{4\rho_{\alpha^{s}}}}lS \\ \nonumber
    c_{\beta}^{s} (\hat{k}_c) = \sqrt{\frac{2J_4 + J_1}{4\rho_{\beta^{s}}}}lS,
\end{eqnarray}
Now since $\rho_{\beta^{s}} = 2\rho_{\alpha^{s}}$, the relation between the velocity of the two types of modes is $c_{\alpha}^s(\hat{k}_c) = \sqrt{2}c_{\beta}^s (\hat{k}_c)$.
\subsubsection{Anisotropy Gaps}
\noindent
The anisotropy terms in Eq.~\ref{eq:HCEF} are a DM interaction ($D$) and an easy axis anisotropy ($K>0$). The easy axis causes a deviation from the 120$^\degree$ order and a gap for the $\alpha_{0}^{s}$ mode. It also splits the otherwise degenerate $\bm{\beta}^{s}$ modes. The DM interaction gaps out the $\bm{\beta}^{s}$ modes that are associated with out of plane spin components (Fig.~\ref{fig:lattice_modes}(b)). The energy gaps as functions of the hamiltonian anisotropy parameters are:
\begin{eqnarray}\label{gaps3}
    E_{\alpha}~ &=& \sqrt{\frac{1}{\rho_{\alpha^{s}}}\left( \frac{3 K^{3}}{J_{eff}^{2}} \right)}S, \\ \nonumber
    E_{\beta_y} &=& \sqrt{\frac{1}{\rho_{\beta^{s}}}\left(2\left(\sqrt{3}D + \frac{K}{2}\right) + \frac{K}{6 J_{eff}}(4\sqrt{3}D - K)\right)}S,\\ \nonumber
    E_{\beta_x} &=& \sqrt{\frac{1}{\rho_{\beta^{s}}}\left( 2\left(\sqrt{3}D + \frac{K}{2}\right) - \frac{K}{6 J_{eff}}(4\sqrt{3}D - K)\right)}S,
\end{eqnarray}
where $J_{eff} = J_2 + J_1$.
\section{Discussion}
\subsection{Nature of the $\mathbf{Q}$~=~0 spin excitations in \ce{Mn3Ge}}\label{origin}
\begin{figure}[t]
    \includegraphics[width=\columnwidth]{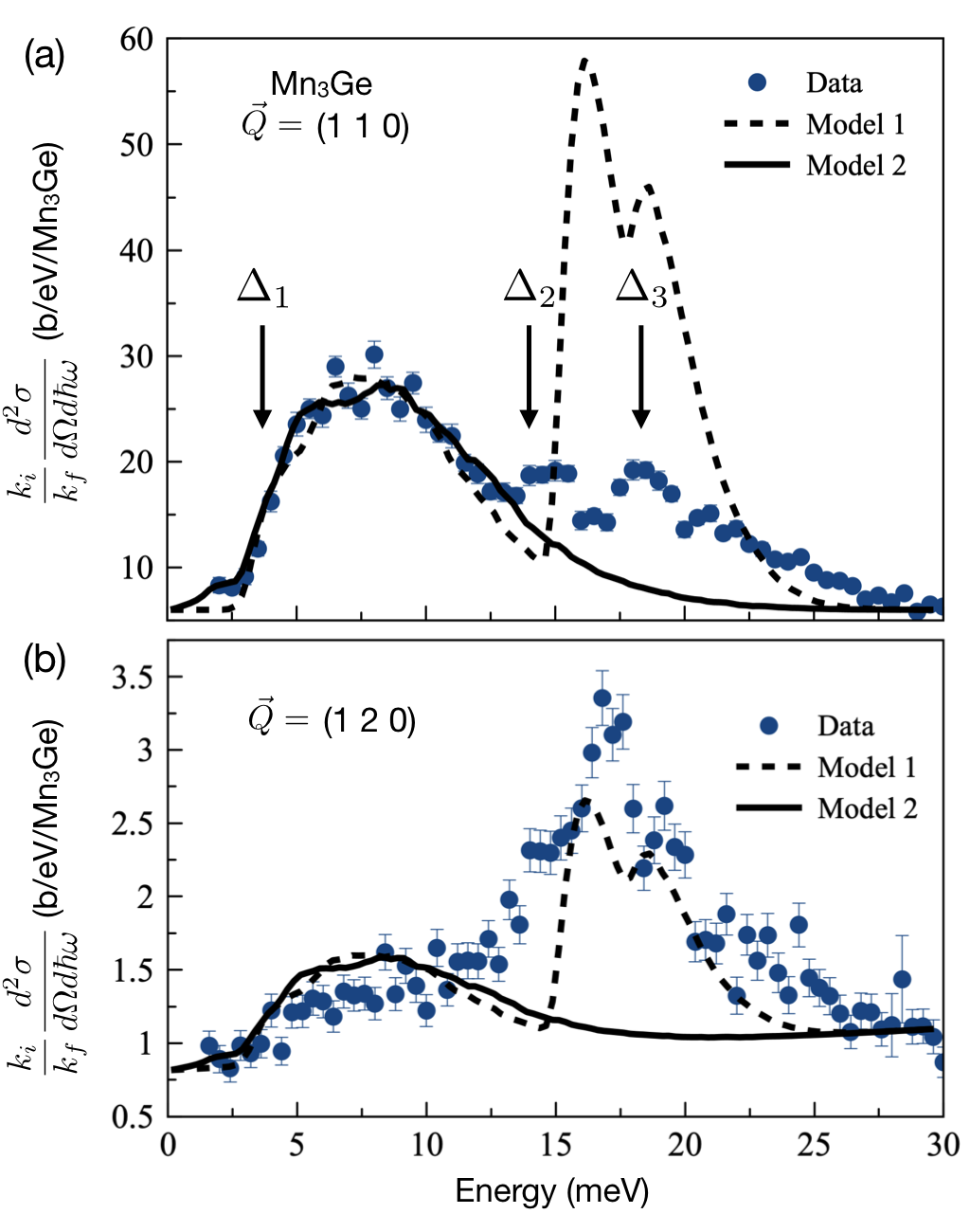}
    \centering
    \caption{(a) The excitation spectrum of \ce{Mn3Ge} at $\mathbf{Q}$~=~(110) compared to the calculated spin-wave spectrum based on two spin Hamiltonian modes including the effects of the instrumental resolution. Model 1 has the  $\alpha$, $\beta_x$ and $\beta_y$ modes located near $\Delta_1$, $\Delta_2$ and $\Delta_3$ respectively. Model 2 has the $\alpha$ mode gapless, while $\beta_x$ and $\beta_y$ are degenerate and gapped by $\Delta_1$. (b) The excitation spectrum at $\mathbf{Q}$~=~(120) where phonon contributions to the scattering intensity dominate. The scattering data were averaged over perpendicular directions of momentum transfer covering $\pm$0.15$~\angstrom^{-1}$ in (a) and over $\pm$0.4$~\angstrom^{-1}$ in (b). The error bars within both panels correspond to 1 standard deviation. }
    \label{fig:model}
\end{figure}

The magnetic excitation spectrum at the $\Gamma$ point ($\mathbf{Q}$~=~0) is shown in Fig.~\ref{fig:model}. Three characteristic energies $\Delta_1, \Delta_2,$ and $\Delta_3$ are indicated. At first glance, it seems natural to associate the $\alpha_0$ mode with $\Delta_1$, and the $\beta_x$ and $\beta_y$ modes with $\Delta_2$ and $\Delta_3$ respectively. It is indeed possible to select parameters in the spin hamiltonian~Eq.~\ref{eq:HCEF} that reproduces the gaps and velocities for each of these modes. However, the comparison of the corresponding resolution convoluted spectrum of neutron scattering with the experimental data in Fig.~\ref{fig:model}(a), reveals clear discrepancies. Specifically, the calculated scattered intensities of the two beta modes are greatly overestimated (dashed line in Fig.~\ref{fig:model}(a)). The relative intensities of these excitations do not depend on details of the exchange interactions introduced, so we conclude this model is not appropriate for \ce{Mn3Ge}.

Instead, we attribute the intensity maxima at $\Delta_2$ and $\Delta_3$ to hybridization of the spin wave excitation with optical phonons that are present in this energy range (Fig.~\ref{fig: lowEspectra}(g)). The vibrational nature of both of these features is confirmed by the neutron scattering spectrum at ${\bf Q}=(120)$ which has the same composition of in- to out-of- plane spin polarization. Fig.~\ref{fig:model}(b) shows generally much less scattering, which is consistent with suppression of the scattering relative to Fig.~\ref{fig:model}(a). However, the intensity in the energy range near $\Delta_2$ and $\Delta_3$ increases and is now stronger than near the $\Delta_1$ mode. Since phonon scattering generally increases with $|{\bf Q}^2|$ this is consistent with hybridized spin-phonon excitation at $\Delta_2$ and $\Delta_3$. This scenario was previously proposed for \ce{Mn3Ge}~\cite{PhysRevB.99.214445}, and is supported by our data. Magneto-elastic coupling effects have been observed in other non-collinear triangular magnets with competing exchange interactions such as multiferroic $RE$MnO$_3$~\cite{ISI:000231310900063,oh2016spontaneous,Holm2018} and $\rm Ni_3V_2O_8$~\cite{ISI:000231310900062}.

With this interpretation the in-plane polarized $\alpha$ mode is gapless, which is consistent with the easy plane nature of the magnetization data~\cite{nayak2016large,OGASAWARA20197}. As described in ref~\cite{liu2017anomalous}, $K$ can be estimated from the in-plane magnetization of \ce{Mn3Ge}, which is directly related to the energy gap of the $\alpha_0$ mode (see Eq.~(\ref{gaps3})). Using this procedure, a spin wave gap of about 0.1~meV$<<\Delta_1$ is estimated. The $\Delta_1$ mode is instead associated with the energy gap for the $\beta_x$ and $\beta_y$ modes. Our field theory indeed predicts these to be degenerate when the $\alpha$ mode is gapless ($K<<J$) (see Eq.~(\ref{gaps3})).

The resolution convoluted $\mathbf{Q}$~=~0 magnetic excitation spectrum can be computed based on a spin hamiltonian (Eq.~\ref{eq:HCEF}) with a gapless $\alpha_0$ mode and the two $\beta$ modes at $\Delta_1$ (see Fig.~\ref{fig:model}(a)). The agreement of this calculation with the observed spectrum of \ce{Mn3Ge} is excellent aside for scattering at $\Delta_2$ and $\Delta_3$ that we associate with magnon-phonon hybridization, which is not in the model. The lack of significant inelastic neutron scattering at low energies associated with the gapless alpha mode is a result of the large spin wave velocity that pushes spectral density to higher energies and reduces the low energy density of states. Similar effects were observed for other magnets with fast spin waves when probed with relatively coarse momentum resolution~\cite{PhysRevB.24.1312,Zaliznyak2004}.

\subsection{Effective spin Hamiltonian for \ce{Mn3Ge}}\label{HamiltonianMnGe}
A general feature of the nearest neighbor JDK model (eq.~\ref{eq:HCEF}) on the kagome lattice is the dispersionless nature of the $\beta_x$ mode~\cite{Harris1992,Chernyshev2015}. This so-called "weather-vane" mode was for example observed in the jarosites kagome systems~\cite{Matan2006,Coomer2006}, which have been described using the JDK model. The weathervane mode can be made to disperse through the introduction of longer ranged exchange interactions. However, there is insufficient information in the broad magnetic excitations that we detect in Mn$_3$Ge to uniquely determine multiple further neighbor interactions. Here the field theory provides important guidance through analytical expressions for the in-plane and out-of-plane velocities and the excitation gaps of the three long-wavelength normal modes (see Eqs.~(\ref{eq.velocities_total},\ref{eq.outofplanevelocities},\ref{gaps3})). 

From the field theory, we find that a model with $J_1$, $J_2$ and $J_3$ cannot describe the dispersive spin excitations of \ce{Mn3Ge}, because both $J_1$ and $J_3$ leave the weathervane mode dispersionless for wave vectors along the $(h00)$ direction. This is inconsistent with the experimental data (Fig.~\ref{fig: lowEspectra}(b,d)). However, a ferromagnetic $J_4$ interaction causes dispersion of the weathervane mode for all directions in reciprocal space. Thus a minimal model for \ce{Mn3Ge} must include $J_1, J_2$, and $J_4$ plus the nearest in-plane neighbor DM interaction of the form ${\bf D}=D\uvec{z}$ to gap the $\beta$ modes.

In the long-wavelength limit the dispersion relation of the soft magnons can be approximated as $E_{i}^{2}(\mathbf{Q})=c_i^2\vert \mathbf{Q}\vert^2 $+$\Delta^2$ with $c_i$ given by Eq.~(\ref{eq.velocities_total}) and Eq.~(\ref{eq.outofplanevelocities}). The energy gap $\Delta_1= 2.9(6)$~meV was determined by fitting a resolution convoluted cross-section based on this dispersion relation to the data in Fig.~\ref{fig:model}(a) for which the absolute intensity scale of our calculation was allowed to vary. We used the spin wave velocities previously determined by fitting constant $\hbar\omega$ slices in section~\ref{sec:SpinPho}. Both the measured velocities and gaps were associated with the $\beta_x$ and $\beta_y$ modes. A least-squares refinement shows a model with the exchange parameters reported in Table.\ref{tab: exchange} can reproduce the measured spin wave velocities. Also reported in Table.\ref{tab: exchange} is the strength of the DM interaction that was obtained from $\Delta_1$ and the refined exchange constants through Eq.~(\ref{gaps3}). For a JDK model with $K\approx0$ this is the only set of values for $J_1S^2$, $J_2S^2$, and $J_4S^2$ that can reproduce the long-wavelength dispersion relation. We note that a relatively strong out-of-plane $J_4$ interaction was refined, which confirms that the magnetism in \ce{Mn3Ge} is 3D.

\begin{table}[]
\begin{tabular}{|c|c|c|c|c|c|}
\hline
                                                                   & $J_1S^2$ & $J_2S^2$ & $J_4S^2$ & $DS^2$       & $KS^2$                          \\ \hline
\begin{tabular}[c]{@{}c@{}}refined value\\      (meV)\end{tabular} & 0(6)     & 34(7)    & $-17(5)$   & 0.02(1) & $\leq$0.01 \\ \hline
\end{tabular}
\caption {Microscopic parameters of the spin Hamiltonian refined in our work for \ce{Mn3Ge} (Eq.~\ref{eq:HCEF} ). A positive (negative) sign for the exchange parameters corresponds to AFM (FM) interactions. Note that $J_1$ and $J_4$ are inter-plane interactions (see Fig.\ref{fig:interplane-interaction}), while $J_2$, D and K are intra-plane interactions.}
\label{tab: exchange}
\end{table}

The $\mathbf{Q}$~=~0 neutron inelastic spectrum of the proposed hamiltonian can be computed with linear spin-wave theory as implemented in spinW~\cite{toth2015linear}. The calculated spectrum was averaged over the six magnetic domains of \ce{Mn3Ge} and convoluted with the 4D resolution function appropriate for the configuration of SEQUOIA that we utilized. The resulting fit is shown in Fig.~\ref{fig:model}(a). The excellent agreement between the calculation and the scattering profile of the $\Delta_1$ mode validates our minimal spin hamiltonian model to describe the long-wavelength low energy magnetism of \ce{Mn3Ge}. While the non-collinear nature of the magnetic order allows magnon decay~\cite{PhysRevB.98.184403} the lack of collective resonant modes beyond those at the $\Gamma$ point is akin to itinerant magnets such as $\rm Mn_3Si$~\cite{PhysRevB.36.2181}. Also distinguishing $\rm Mn_3Ge$ from local moment magnetism, the ordered moment $M_\chi=2.2(1)~\mu_B$ is considerably reduced from the full moment expected for known oxidation states of Mn ions such as Mn$^{2+}$ (5.92$~\mu_B$) and Mn$^{4+}$ (3.87$~\mu_B$). 

\section{Conclusion}\label{conclusion}
We have shown $\rm Mn_3Ge$ undergoes a second-order magnetic phase transition at $T_N=365$~K, the order parameter of which is an anti-chiral triangular spin structure superimposed with weak in-plane ferromagnetism described by a single two-dimensional irreducible representation $\Gamma_9$. Probed here by inelastic neutron scattering, the magnetic excitation spectrum features long wavelength spin waves and a broad continuum centered at 75~meV. We construct a field theory of antiferromagnetic triangular simplexes to classify the three Goldstone modes as an in-plane polarized gapless mode and a doublet of out of plane polarized modes. The scattering data also provide evidence for a pair of magneto-elastic modes near 20~meV that are enabled by the non-collinear nature of the magnetic order. The selection of the $\Gamma_9$ spin structure, the Goldstone modes, and the magnetic response to an applied field can be described by a minimal spin Hamiltonian ${\cal H}_{JDK}$ consisting of two inter-plane Heisenberg interactions $J_1$ (weak) - $J_4$ (strong, FM), intra-plane Heisenberg $J_2$ (strong AFM) and weak anti-chiral Dzyaloshinskii-Moriya interactions $D$, and an even weaker single-ion anisotropy term $K$ favoring spin orientations that bisect two triangular simplexes. Field theory links experimental observables such as the spin wave velocities and their excitation gaps to the parameters of the spin Hamiltonian, facilitating their controlled determination. Interactions between the collective magnetism and itinerant electrons are apparent in the strong damping of the spin waves, the reduced local moment size, the range of magnetic interactions, and of course the anomalous transport properties, all of which ${\cal H}_{JDK}$ can be the basis for modeling and understanding.

\begin{acknowledgments}
We greatly appreciate the technical support from T.~Dax, R.~Erwin, S. ~Shannon and M.~T.~Hassan at the NIST Center for Neutron Research. We thank Shu Zhang for illuminating discussions and patient hearing of ideas. This work was supported as part of the Institute for Quantum Matter, an Energy Frontier Research Center funded by the U.S. Department of Energy, Office of Science, Basic Energy Sciences under Award No. DE-SC0019331. Y.C. and J.G. contributed equally to this work. J.G. acknowledges support from the NSERC Postdoctoral Fellowship Program. C.B. was supported by the Gordon and Betty Moore Foundation through the EPIQS program GBMF-4532. A portion of this research used resources at the High Flux Isotope Reactor and Spallation Neutron Source, a DOE Office of Science User Facility operated by the Oak Ridge National Laboratory. We also acknowledge the support of the National Institute of Standards and Technology, U.S. Department of Commerce. The identification of any commercial product or trade name does not imply endorsement or recommendation by the National Institute of Standards and Technology. This work is also partially supported by CREST (JPMJCR18T3), Japan Science and Technology Agency (JST), by Grants-in-Aids for Scientific Research on Innovative Areas (15H05882 and 15H05883) from the Ministry of Education, Culture, Sports, Science, and Technology of Japan, by Grants-in-Aid for Scientific Research (19H00650), and by New Energy and Industrial Technology Development Organization. 
\end{acknowledgments}

\appendix
\counterwithin{figure}{section}
\counterwithin{table}{section}

\section{Polarized neutron diffraction}\label{sec:Appendix11}

\begin{table*}[ht]
%\begin{ruledtabular}
\begin{tabular}{|c|c|c|c|c|c|c|c|c|c|c|c|c|c|c|}
\hline
                         & \multicolumn{8}{c|}{0 T}                                & \multicolumn{6}{c|}{2 T}                  \\
\hline
     & \multicolumn{4}{c|}{Horizontal field }     & \multicolumn{4}{c|}{Vertical field}     & \multicolumn{6}{c|}{Vertical field}                   \\
\hline
$\bf Q$&$\tilde{\sigma}^{obs}_{++}$&$\tilde{\sigma}^{cal}_{++}$& $\tilde{\sigma}^{obs}_{-+}$ & $\tilde{\sigma}^{cal}_{-+}$ & $\tilde{\sigma}^{obs}_{++}$ & $\tilde{\sigma}^{cal}_{++}$ & $\tilde{\sigma}^{obs}_{-+}$ & $\tilde{\sigma}^{cal}_{-+}$ & $\tilde{\sigma}^{obs}_{++}$ & $\tilde{\sigma}^{cal}_{++}$ & $\tilde{\sigma}^{obs}_{-+}$ & $\tilde{\sigma}^{cal}_{-+}$ & $\tilde{\sigma}^{obs}_{--}$ & $\tilde{\sigma}^{cal}_{--}$\\ \hline
(001)$^*$&0.002&0.000&0.001&  0.000      &  0.002      &   0.000     &  0.000     &  0.000      &   -     &    -     &    -     &    -     &   -    &  -     \\ \hline
 (002) &   0.165     &  0.153      &   0.014     &    0.016    &   0.182     &    0.157    &   0.004    &   0.006     &   0.156     &    0.155     &    0.019     &   0.016      &   0.148    &    0.155   \\ \hline
 (100) & 0.197      &  0.176      &  0.121      &   0.120     &   0.314     &   0.274     &  0.003     &   0.000     &  0.009      &   0.005      &    0.021     &   0.000      &  0.506     &  0.418     \\ \hline
 (101)$^*$ & 0.334    &  0.369      &   0.279     &   0.270     &   0.514     &   0.568     &   0.108    &   0.147     &   0.040     &    0.062     &    0.103     &    0.124     &  0.719     &   0.811    \\ \hline
 (102) &    0.234    &  0.236      &    0.103    &  0.110  &    0.387    &  0.294      &0.039   &  0.044      &  0.085       &  0.100       &   0.046      &    0.007     &  0.479  &  0.454     \\ \hline
 (200)$^*$& 0.059        &   0.057     &    0.005    &   0.002     &   0.055     &   0.047     &   0.004    &   0.000     &   0.065     &    0.053     &   0.014      &    0.000     &    0.061   &   0.053    \\ \hline
 (201)$^*$&    0.149    &   0.128     &   0.011     &   0.007     &   0.167     &   0.124     &  0.005     &  0.001      &   0.161     &    0.128     &     0.032    &  0.000       &  0.138     &   0.128    \\ \hline
 (202) &   0.037     &   0.059     &  0.011      &  0.001      &   0.050     &   0.047     &   0.003    &   0.000     &   -     &      -   &    -     &   -      &  -     &   -    \\ \hline
(300)  &  0.502      &   0.507     &   0.142     &   0.124     &   0.729     &   0.678     &   0.014    &   0.000     &   -     &    -     &     -    &   -      &  -     &  -     \\ \hline
(301)$^*$&    0.000    &   0.000     &   0.000     &    0.000    &   0.011     &   0.000     &    0.000   &    0.000    &  -      &    -     &    -     &    -     &    -   &  -     \\ \hline\hline
 (001)$^{*}$ &  0.038      &  0.000      &   0.028     &  0.000      &  0.021      &   0.000     &   0.010    &  0.000      &   -     &     -    &   -      &     -    &    -   &   -    \\ \hline
 (002) &    0.350    &   0.418     &   0.083     &    0.026    &   0.213     &    0.274    &   0.030    &   0.000     &   0.178     &   0.259      &     0.075    &    0.017     &   0.155    &    0.259   \\ \hline
 (110) &   2.081     &   2.066     &    0.772    &  0.773      &   1.450     &   1.466     &   0.070    &  0.000      &  6.071      &   6.219      & 0.688         &   0.000      &  0.407     &   0.324    \\ \hline
 (111) &   0.004     &   0.000     &   0.000     &   0.000     &   0.000     &   0.000     &   0.000    &   0.000     &    -    &    -     &   -      & -        &    -   &   -    \\ \hline
 (112) &   2.169     &   2.231     &    0.674    &  0.541    &   1.535     &   1.461     &   0.328    &   0.157     &     5.862   &    4.646     &            0.258     &   0.064    &   0.202 & 0.943    \\ \hline
\end{tabular}
%\end{ruledtabular}
\caption{The measured ($\tilde{\sigma}^{obs}$) and best fit ($\tilde{\sigma}^{cal}$)  extinction encumbered polarized Bragg diffraction cross-sections in units of barn/formula unit for Mn$_3$Ge in low field and at 2 Tesla. The upper part of the table present peaks collected in the $(h0\ell)$ plane while the lower part of the table present peaks collected in the $(hh\ell)$ plane. $+$ and $-$ indicate the neutron polarization direction before and after the sample relative to the guide field. The relative statistical errors on the individual values of $\tilde{\sigma}^{obs}$ determined by Gaussian least-squares fitting of rocking curves are less than 1\%. The reduced $\chi^2$ for the overall structural fits in zero field are 5264(10219) and 4465(6566) for the ($h0\ell$) and ($hh\ell$) planes respectively for $B\approx 0$ ($B=2$~T). This  indicates systematic errors exceed statistical errors. Peaks that may be affected by multiple scattering are marked by $*$ with the associated intermediate reciprocal lattice point(s) listed in Table~\ref{tab:muti_scattering}.}
\label{tab:structurefactors}
\end{table*}

\subsection{From count rate to cross section}

The $\bf Q$ dependence of the intensity near a Bragg peak $\bf G$, is a measure of the instrumental momentum resolution and is given by
\begin{equation}
    I({\bf Q})= \tilde{\sigma}_{\bf G}N{\cal C}R_{\bf G}\exp{[-\frac{1}{2}({\bf G}-{\bf Q})^{\bf{T}}\bf{M}_{\bf G}({\bf G}-{\bf Q})}]
\end{equation}
Here $\tilde{\sigma}_{\bf G}$ is the total $\bf Q$ integrated Bragg intensity, $\bf{M}_{\bf G}$ is the resolution matrix,  $R_{\bf G}$ ensures normalization of the resolution function, and $N{\cal C}$ is an overall instrument normalization factor~\cite{Chesser:a09670}. Denoting the rocking angle integrated intensity by $A({\bf G})$ and $\hat{\bf y}$ the Cartesian coordinate direction corresponding to the trajectory of the rocking scan perpendicular to ${\bf G}$, we have
\begin{equation}
    \tilde{\sigma}_{\bf G}=\frac{A({\bf G})|{\bf G}|}{N{\cal C}R_{\bf G}}\sqrt{({\bf M}_{\bf G})_{yy}/2\pi}  
\label{sigma}
\end{equation}
Table~\ref{tab:structurefactors} shows the Bragg diffraction cross sections obtained from polarized beam rocking scans scans based on Eq.~\ref{sigma}. 

\begin{table}[ht]
    \centering
    %\begin{ruledtabular}
    
    \begin{tabular}{|c|c|c|}
    \hline
    Zone&Target $\bf G$&Intermediate ${\bf G}^\prime$\\ \hline
    $(h0\ell)$&$(001)$&$(302)$\\ %\hline
    $(h0\ell)$&$(101)$&$(02\bar{1})(2\bar{2}\bar{1})$\\ %\hline
    $(h0\ell)$&$(200)$&$(010)(1\bar{1}0)(110)(2\bar{1}0)$\\ %\hline
    $(h0\ell)$&$(201)$&$(120)(3\bar{2}0)(31\bar{1})(4\bar{1}\bar{1})$\\ %\hline
    $(h0\ell)$&$(301)$&$(30\bar{2})$\\ %\hline%\hline
    $(hh\ell)$&$(001)$&$(10\bar{1})(102)(01\bar{1})(012)$\\ \hline
    \end{tabular}
    
    %\end{ruledtabular}
    \caption{The target Bragg reflection ${\bf G}={\bf k}_i-{\bf k}_f$ and intermediate Bragg points ${\bf G}^\prime$ where $||{\bf k}_i-{\bf G}^\prime|-k_i|<\delta k_i$ so that multiple Bragg diffraction is possible. Specifically the following diffraction processes: ${\bf k}_i\rightarrow{\bf k}_i-{\bf G}^\prime\rightarrow{\bf k}_f$ and ${\bf k}_i\rightarrow{\bf k}_f\rightarrow{\bf k}_f+{\bf G}^\prime-{\bf G}$. Here $k_i=2.6636$~\AA$^{-1}$ corresponding to $E_i=14.7$~meV and $\delta k_i=0.02$~\AA$^{-1}$ is set by the energy resolution of the instrument.}
    \label{tab:muti_scattering}
\end{table}

\subsection{Extinction}

To account for secondary extinction, the actually measured cross-sections $\tilde{\sigma}$ were related to the theoretical Born limit cross-sections $\sigma$ (discussed in section~\ref{crosssection}) by the empirical extinction formula used by Fullprof ~\cite{Rodriguez1990}
\begin{equation}
    \tilde{\sigma}={\sigma}/{\sqrt{1+\frac{10^{-3}y\lambda^{3}}{\sin{2\theta}}\sigma}}
\end{equation}
where $\lambda$ is the neutron wavelength, $2\theta$ is the scattering angle and $y$ is the extinction parameter. Within the refinement, the extinction and normalization parameters were constrained to be the same for peaks that are measured in the same configuration of the sample and instrument.

\subsection{polarization analysis}
In a fully polarized neutron scattering experiment, the cross-section is resolved into four channels denoted ($++,--,+-,-+$). Here $+$ and $-$ indicate neutron spins parallel and anti-parallel to the guide field, respectively. The first sign gives the polarization direction of the incident beam and the second sign indicates the polarization vector of the neutrons received by the detector. The guide field direction was either parallel to momentum transfer $\bf{Q}$ (HF) or perpendicular to the scattering plane (VF).

For low field data, the cross-sections of the non-spin-flip (NSF) $++$ and spin-flip (SF) $-+$ channels were collected. For the neutron diffraction experiments in a 2 T magnetic field, the cross-sections of all four channels ($++$,$-+,+-,--$) were measured. The time dependence of the transmission and polarization of the $^3$He cells were characterized by measuring the flipping ratio of the nuclear Bragg diffraction from a pyrolytic graphite (PG) sample before and after the $\rm Mn_3Ge$ experiment. The inferred time dependence of the polarization characteristics of the instrument was verified by measuring the flipping ratio of Bragg peaks associated with aluminum in our sample mount multiple times during the experiment. A time-dependent correction was applied to the polarized beam diffraction data. Here the depolarizing effects of the sample were neglected due to the high flipping ratio of the zero field experiment and the small magnetization of $\ce{Mn3Ge}$. After polarization correction, the integrated intensity of each Bragg peak in each polarization channel was obtained by fitting the rocking curves to Gaussian functions.

\subsection{Polarized neutron cross-sections}
\label{crosssection}
For each model described by one IR of $G_{0}$, the following formulas are used to calculate the cross-sections in different channels based on the nuclear structure factor $F_{N}$ and the perpendicular to $\bf Q$ projection of the magnetic vector structure factor ${\bf F}_{M}^{\perp}=\hat{\bf Q}\times {\bf F}_{M}\times\hat{\bf Q}$:
\begin{eqnarray}
\sigma_{++}&=&|F_{N}+\hat{\bf p}\cdot{\bf F}_{M}^\perp|^{2}\label{nsfp}\\
\sigma_{--}&=&|F_{N}-\hat{\bf p}\cdot{\bf F}_{M}^\perp|^{2}\label{nsfm}\\
\sigma_{+-}&=&|{\bf F}_{M}^\perp|^{2}-|\hat{\bf p}\cdot {\bf F}_{M}^\perp|^{2}-i\cdot \hat{\bf p}\cdot({\bf F}_{M}^\perp\times {\bf F}_{M}^{\perp*})\\
\sigma_{-+}&=&|{\bf F}_{M}^\perp|^{2}-|\hat{\bf p}\cdot {\bf F}_{M}^\perp|^{2}+i\cdot \hat{\bf p}\cdot({\bf F}_{M}^\perp\times {\bf F}_{M}^{\perp *})\label{pol}.
\end{eqnarray}
Here $\hat{\bf p}$ is the unit vector indicating the polarization direction (guide field). In our refinement, the magnetic form factor of \ce{Mn^{2+}} is included in ${\bf F}_{M}$~\cite{tomiyoshi1983triangular}.

Table~\ref{tab:structurefactors} provides the measured and fitted polarized Bragg diffraction cross-sections for $\rm Mn_3Ge$ that are the basis for Fig. 1(e,f). Table~\ref{tab:muti_scattering} reports the Bragg peaks that may be impacted by multiple Bragg diffraction.

\subsection{Magnetic structure refinement}

\begin{table*}[ht]
%\begin{ruledtabular}
\begin{tabular}{|c|c|c|c|c|c|c|c|c|c|c|c|c|}
\hline
IR&e&\makecell{$\{~6^{+}_{001}~|~t~\}$\\$\{~6^{-}_{001}~|~t~\}$}&\makecell{$3^{+}_{001}$\\$3^{-}_{001}$}&$\{~2_{001}~|~t~\}$&\makecell{$2_{010}$\\$2_{100}$\\$2_{110}$}&\makecell{$\{~2_{1-10}~|~t~\}$\\$\{~2_{210}~|~t~\}$\\$\{~2_{120}~|~t~\}$}&$i$&\makecell{$\{~-6^{+}_{001}~|~t~\}$\\$\{~-6^{-}_{001}~|~t~\}$}&\makecell{$-3^{+}_{001}$\\$-3^{-}_{001}$}&$\{~{\rm m}_{001}~|~t~\}$&\makecell{$\rm{m}_{010}$\\$\rm{m}_{100}$\\$\rm{m}_{110}$}&\makecell{$\{~{\rm m}_{1-10}~|~t~\}$\\$\{~{\rm m}_{210}~|~t~\}$\\$\{~{\rm m}_{120}~|~t~\}$}\\
\hline
$\Gamma_{2}$(A$_{1u}$)&1&1&1&1&1&1&-1&-1&-1&-1&-1&-1\\
$\Gamma_{3}$(A$_{2u}$)&1&1&1&1&-1&-1&1&1&1&1&-1&-1\\
$\Gamma_{4}$(A$_{2g}$)&1&1&1&1&-1&-1&-1&-1&-1&-1&1&1\\
$\Gamma_{5}$(B$_{1g}$)&1&-1&1&-1&1&-1&1&-1&1&-1&1&-1\\
$\Gamma_{6}$(B$_{1u}$)&1&-1&1&-1&1&-1&-1&1&-1&1&-1&1\\
$\Gamma_{7}$(B$_{2g}$)&1&-1&1&-1&-1&1&1&-1&1&-1&-1&1\\
$\Gamma_{9}$(E$_{1g}$)&2&1&-1&-2&0&0&2&1&-1&-2&0&0\\
$\Gamma_{10}$(E$_{1u}$)&2&1&-1&-2&0&0&-2&-1&1&2&0&0\\
$\Gamma_{11}$(E$_{2g}$)&2&-1&-1&2&0&0&2&-1&-1&2&0&0\\
$\Gamma_{12}$(E$_{2u}$)&2&-1&-1&2&0&0&-2&1&1&-2&0&0\\\hline
\end{tabular}
%\end{ruledtabular}
\caption{Character table of conjugate classes for different irreducible representations of the 'little group' $G_{0}$ that keeps the wave vector $\mathbf{k}$~=~$\mathbf{0}$ unchanged. $t=[0~0~1/2]$ and the subscript (1 or 2) of the Mulliken symbols are determined by $C_{2}'$ axis $[100]$}
\label{tab:chararctertable}
\end{table*}

\begin{figure}[t]
    \includegraphics[width=\columnwidth]{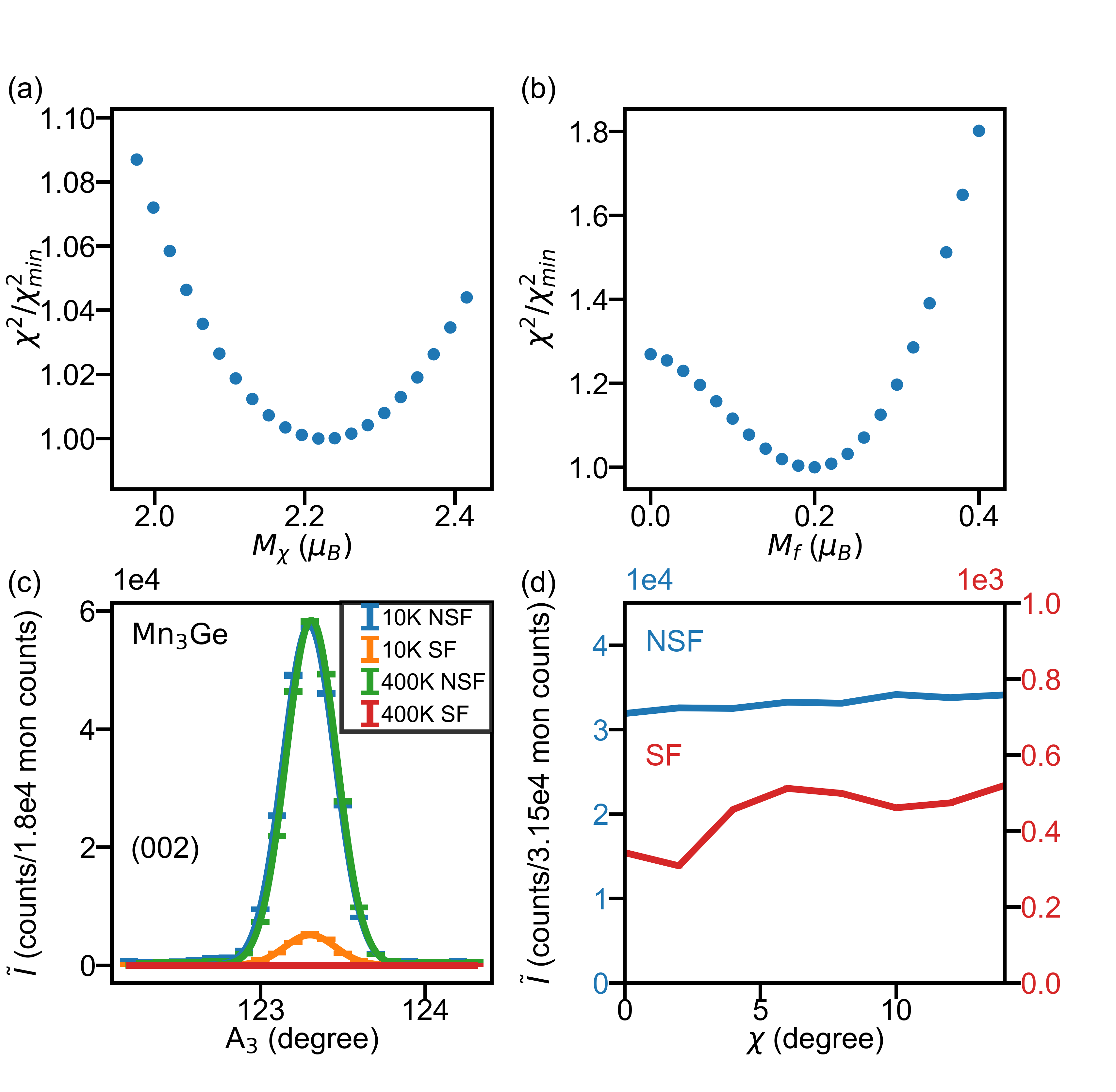}
    \centering
    \caption{Results of refinement for diffraction at B=0 T. (a) $\chi^{2}$ against the anti-chiral component and (b) the ferromagnetic moment.(c) Rocking scans of (002) in and out of the magnetic phase with the HF configuration, the non-spin-flip scattering is also displayed for reference. (d) The $\chi$ scan for (002) performed at room temperature indicates the spin-flip scattering is not produced by multiple scattering.}
    \label{fig:chi_square}
\end{figure}

The  ${\bf k}={\bf 0}$ magnetic ordering implies the magnetic Bragg peaks of \ce{Mn3Ge} coincide in $\bf Q$-space with structural Bragg peaks. Table~\ref{tab:chararctertable} gives the characters of classes for different IRs of the `little group' $G_{0}$ that keeps the ordering wave vector $k=0$ unchanged. $\Gamma_{9}$ is the only IR that is consistent with the observed diffraction pattern of \ce{Mn3Ge}. Fig.~\ref{fig:chi_square}(a,b) shows the $\chi^{2}$ dependence of the magnetic refinement as a function of the moment size of the anti-chiral ($M_{\chi}$) and ferromagnetic components ($M_{f}$). Table~\ref{tab:structurefactors} compares the measured and calculated diffraction polarized beam cross sections. Peaks marked with * and acquired in the $(h0l)$ are potentially impacted by multiple scattering through the intermediate Bragg points indicated in Table~\ref{tab:muti_scattering}. The size of the ferromagnetic moment $M_{f}=0.2(1)\mu_{B}$ is directly related to the magnitude of the spin-flip scattering at (002) (Fig.~\ref{fig:chi_square} (c)). We performed a $\chi$ rotation scan for the (002) reflection to rule out the possibility of multiple scattering as a source of  spin-flip scattering at (002). The $\chi$ scan is shown in Fig.~\ref{fig:chi_square}(d) and both the SF and NSF scattering at (002) are constant to within alignment tolerances over the full range of 0 $\leq$ $\chi$ $\leq$ 15$\degree$. This observation excludes multiple scattering as a source of Bragg scattering at (002).

\section{Ferromagnetic moment of \ce{Mn3Ge}}\label{sec:Appendix12}
\begin{figure}[t]
    \includegraphics[width=\columnwidth]{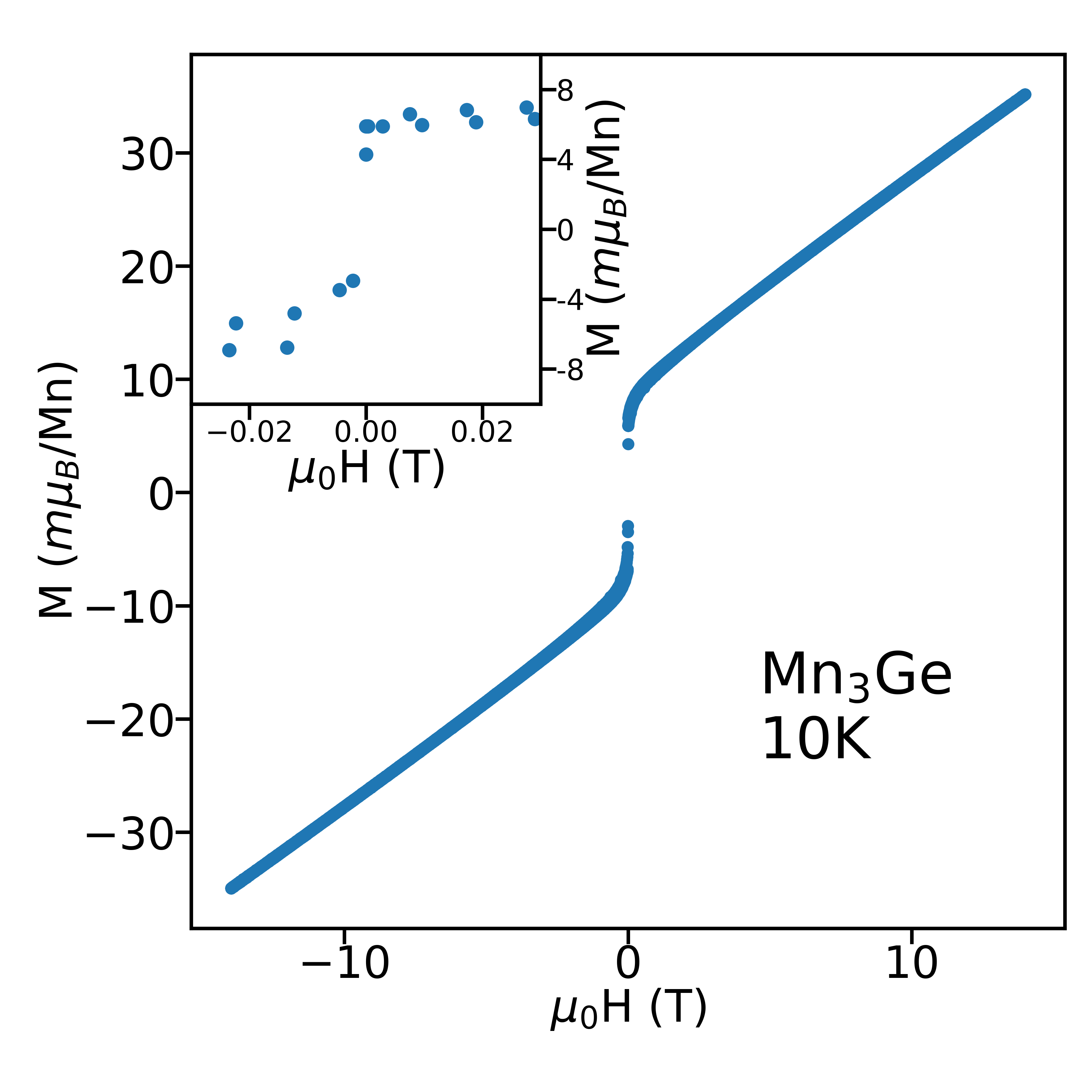}
    \centering
    \caption{Bulk magnetization of the \ce{Mn3Ge} single crystal that was used for the diffraction experiments reported in this paper. The data were collected at $T$~=~10~K and the field was applied along the [100] direction. The inset details the low field regime showing a zero-field magnetization of about 0.007$\mu_B$/Mn.}
    \label{fig:MvsH}
\end{figure}

As described in the main text, refinement of the single-crystal polarized beam neutron diffraction data yielded a ferromagnetic moment of 0.2(1)$\mu_B$/Mn for \ce{Mn3Ge}. This is two orders of magnitude greater than the value obtained from bulk magnetization measurements on the exact same \ce{Mn3Ge} single crystal that we used for neutron diffraction. The data is shown in Fig.~\ref{fig:MvsH} and it reveals a basal plane magnetization of 0.007$\mu_B$/Mn which is consistent with previous works~\cite{nayak2016large,Kiyohara2016}.

Below, we offer two possible explanations for the different ferromagnetic moments extracted by neutron diffraction and bulk magnetization measurements.

\subsection{Screening of nm scale ferrimagnetic spin clusters}
The hexagonal $D0_{19}$ structure of \ce{Mn3Ge} was synthesized with excess Mn, annealed at high temperature ($\sim$1000~K), and later quenched cooled to room temperature in water~\cite{doi:10.1063/1.5064697,nayak2016large}. The $D0_{19}$ structure of \ce{Mn3Ge} is not a stable phase at room temperature, and transforms into a tetragonal $D0_{22}$ structure over a long time scale or when heated above 500~K~\cite{doi:10.1143/JPSJ.16.1995}. Tetragonal Mn$_3$Ge is an easy axis ferrimagnet and ordered as grown at $T_c=800~K$~\cite{doi:10.1143/JPSJ.16.1995}. A fully disordered phase can be obtained by annealing at a lower temperature (650~K) and it exhibits the same critical temperature and non-saturated magnetic moment as the $D0_{19}$ phase of \ce{Mn3Ge}~\cite{doi:10.1143/JPSJ.16.1995}. 

\begin{figure}[t]
    \includegraphics[width=\columnwidth]{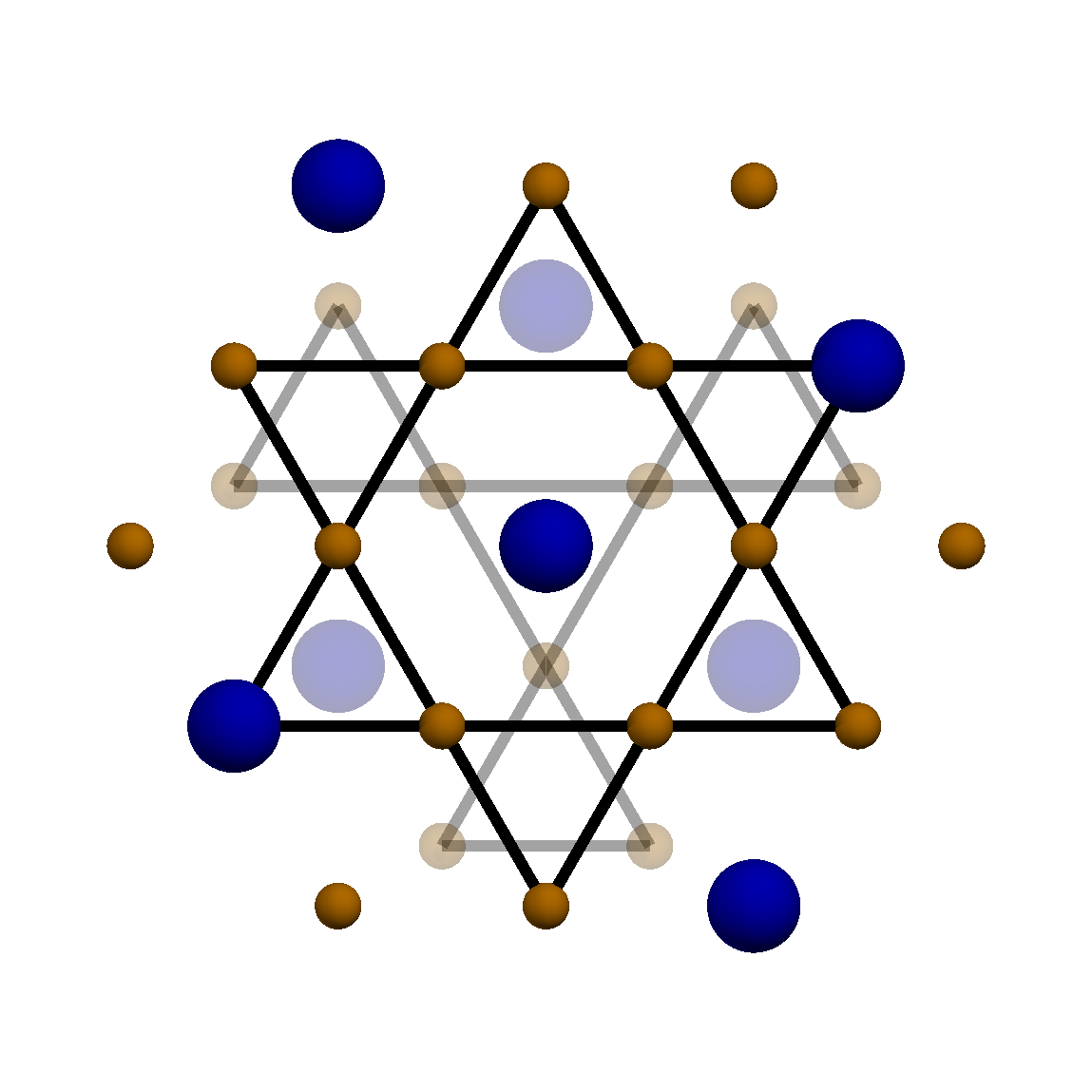}
    \centering
    \caption{Representation of the atomic layers at the boundary of a (112) $D0_{22}$ tetragonal phase of \ce{Mn3Ge} and a (001) $D0_{19}$ hexagonal phase of \ce{Mn3Ge}. The bottom shaded layer is the hexagonal phase, and the dark orange and blue dots represent the Mn and Ge ions respectively.}
    \label{fig:tetragonal}
\end{figure}

Thus, while the amount of the tetragonal phase can be reduced to a few percent in good samples~\cite{nayak2016large}, the tetragonal phase remains and can influence the magnetism of nominally hexagonal $D0_{19}$ \ce{Mn3Ge} samples. For example, as shown in Fig.~\ref{fig:tetragonal}, a slight rearrangement of atoms within the basal plane of the $D0_{19}$ structure leads to an epitaxially embedded $D0_{22}$ phase. Here the (112) direction of the tetragonal $D0_{22}$ phase is parallel to the c-axis of the $D0_{19}$ phase and features the same distance between adjacent "kagome" layers. Hence, the (112) Bragg peak of the $D0_{22}$ phase coincides with the (002) Bragg peak of the $D0_{19}$ phase. 

A tetragonal phase embedded into the hexagonal phase of \ce{Mn3Ge} may explain the discrepancy between the ferromagnetism detected by magnetometry versus neutron diffraction. Assume tetragonal clusters develop a ferrimagnetic moment just as the bulk tetragonal phase. For consistency with the small moment detected by bulk magnetometry, these clusters must be magnetically screened via an antiferromagnetic coupling to the majority $D0_{19}$ structure of \ce{Mn3Ge}. Magnetic spin-flip Bragg scattering could nonetheless remain at (002) if the epitaxial impurity magnetism is coherent at least to the 10~nm length scale. This could occur either if the tetragonal impurity clusters reach this size or if small neighboring clusters become magnetically correlated amongst each other. 

To further elucidate this possibility, we consider measurements of the (002) reflection in a 2~T vertical field that is oriented within the basal plane of \ce{Mn3Ge}. From Eq.~\ref{nsfp}-\ref{pol} at (002) we have:
\begin{eqnarray}\label{eq:002Xsection1}
\sigma_{++}-\sigma_{--}\propto\sum_{i} V_{i}F_{Ni}(\hat{\bf H}\cdot{\bf F}_{Mi}^\perp)\\
\label{eq:002Xsection2}
\sigma_{+-}=\sigma_{-+}\propto\sum_{i}V_{i}|\hat{\bf p}\times{\bf F}_{Mi}^\perp|^{2}
\end{eqnarray}
The summation is over distinct regimes $i$ of the sample large enough to diffract coherently, $V_{i}$ is the volume fraction, $F_{Ni}$ is the nuclear structure factor at (002) and ${\bf F}_{Mi}^\perp$ is the basal plane component of the magnetic vector structure factor at (002). The impurity clusters are associated with dislocation of ions within the basal plane so $F_{Ni}(00L)\equiv F_{N}(00L)$ for integer $L$ including (002). We see that $\sigma_{++}-\sigma_{--}$ is proportional to the magnetization of the sample along the field direction which must vanish if the impurity moments are screened. On the other hand, the spin-flip magnetic Bragg cross-sections are produced by the sum squared magnetization perpendicular to the applied field which is not eliminated by the aforementioned screening as long as it occurs on length scales beyond the $\approx 10$~nm coherence volume. This might be possible for sufficiently large patches of a magnetically anisotropic epitaxial minority phase such as tetragonal \ce{Mn3Ge}~\cite{doi:10.1143/JPSJ.16.1995}. This scenario reconciles the polarized neutron spin-flip cross-sections at (002) with the magnetization data. The screening mechanism is unclear and potentially interesting. This mechanism suggests there might be high field anomalies in the magnetization as the applied field overwhelms the screening of $>10$ nm tetragonal impurity clusters. 

The volume fraction of the tetragonal cluster can be estimated according to Eqs.~\ref{nsfp}-\ref{pol} where the ferromagnetic moment per Mn atom of the tetragonal phase projected in the kagome plane is defined by $m_{c}$ and its volume fraction by $x_{c}$. The corresponding screening moment per Mn atom $m_{s}$ and volume fraction $x_{s}$ should satisfy the following equations based on the experimental observation at \textbf{Q}$=(002)$:
\begin{eqnarray}
\sigma_{++}\textrm{(VF)}-\sigma_{--}\textrm{(VF)}=0=m_{c}x_{c}+m_{s}x_{s}\\
\sigma_{+-}\textrm{(HF)}=\sigma_{-+}\textrm{(HF)}=A(m_{c}^{2}x_{c}+m_{s}^{2}x_{s}).
\end{eqnarray}
Here $A$ is an overall factor $A~=~(3\gamma_{n} r_{e}/2)^{2}$, where 3 is the number of Mn atoms in a formula unit, $\gamma_{n}~=~1.91$ is the gyromagnetic ratio of a neutron and $r_{e}$ is the classical radius of an electron. Solving for the volume fraction yields
\begin{equation}
x_{c}=\frac{\sigma_{+-}(\textrm{HF})}{Am_{c}(m_{c}+m_{s})}.
\end{equation}
In our particular case, the projected moment of the tetragonal phase is about $m_c=0.2~\mu_{B}/\ce{Mn}$~\cite{Qian_2014} and the spin-flip cross-section $\sigma_{+-}(\textrm{HF})=14~\textrm{mbarn}/f.u.$ is listed in Table~\ref{tab:structurefactors}. The volume fraction of the tetragonal phase is then approximately given by $x_{c}=0.1(\mu_B/\ce{Mn})/(m_{s}+0.2\mu_B/\ce{Mn})$. In the thin shell limit, a small cluster volume fraction of few percents is enough to reproduce the observed spin-flip scattering at \textbf{Q}~=~$(002)$, where only a few layers of hexagonal \ce{Mn3Ge} lying close to the cluster act as a screening shell and with $m_{s}$ being on the order of several Bohr magneton per Mn atom.

Recent magnetization measurements on thin film samples of \ce{Mn3X} show low field saturation occurs with a ferromagnetic moment of about $0.2~\mu_{B}/f.u.$ and exhibits exchange bias~\cite{OGASAWARA20197,PhysRevMaterials.2.051001}. These results indicate  coupling of the bulk magnetization of \ce{Mn3Ge} with layered ferromagnetic defects, and they support the scenario described above consisting of antiferromagnetic interactions between tetragonal minority regions in otherwise hexagonal \ce{Mn3Ge}.

\subsection{Screening of local moment ferromagnetism by orbital magnetization} 

Another possibility relies on screening of the Mn moment by orbital magnetization, which cannot be detected by neutron scattering because of the sharply peaked magnetic form factor associated with the orbital magnetization. Suppose the ferromagnetic component of the local moment is weakly coupled to the anti-chiral component, and couples antiferromagnetically to an extended orbital moment. At low fields, there will be magnetic Bragg scattering from local moments at (002) because the compensating orbital moment remains undetected due to its unfavorable form factor. At higher fields, a spin-flop transition may occur turning both the ferromagnetic and the orbital moment perpendicular to the applied field. Again because neutrons only detect the local moments there would be spin-flip magnetic Bragg scattering in the vertical field configuration at (002). 

\noindent
\bibliography{MnGe}
\end{document}